\documentclass[numberedappendix]{emulateapj}
\usepackage{amsmath}
\usepackage{apjfonts}
\usepackage{color}
\usepackage{verbatim}
\usepackage[utf8]{inputenc}
\graphicspath{{figures/}}

\providecommand{\sun}{\ensuremath{\mathord\odot}}
\newcommand{\msun}[0]{{\text{M}_{\sun}}}
\providecommand{\degr}{\ensuremath{^\circ}}
\newcommand{\cdeg}{\degr}

\DeclareMathOperator*{\Laplacian}{\Delta}

\providecommand{\gtrsim}{\ga}
\providecommand{\lesssim}{\la}

\newcommand{\newcodename}[2]{\newcommand{#1}[1][]{\textsc{#2}##1}}

\newcodename{\maya}{Maya}
\newcodename{\mayakranc}{MayaKranc}
\newcodename{\cactus}{Cactus}
\newcodename{\carpet}{Carpet}
\newcodename{\kranc}{Kranc}
\newcodename{\whisky}{Whisky}
\newcodename{\twopunctures}{2Punctures}
\newcodename{\ahfinderdirect}{AHFinderDirect}

\newcommand{\newacronym}[3]{\newcommand{#1}[1]{#3##1 (#2##1)\renewcommand{#1}[1]{#2####1}}}

\newacronym{\NSF}{NSF}{National Science Foundation}
\newacronym{\NASA}{NASA}{National Aeronautics and Space Administration}
\newacronym{\lisa}{LISA}{the Laser Interferometer Space Antenna}
\newacronym{\ligo}{LIGO}{Laser Interferometer Gravitational-wave Observatory} 
\newacronym{\Caltech}{Caltech}{California Institute of Technology}
\newacronym{\MIT}{MIT}{Massachusetts Institute of Technology}
\newacronym{\sph}{SPH}{smooth particle hydrodynamics}
\newacronym{\tsi}{TSI}{the Terascale Supernova Initiative}
\newacronym{\wmap}{WMAP}{the Wilkinson Microwave Anisotropy Probe}
\newacronym{\cmbr}{CMBR}{cosmic microwave background}
\newacronym{\ibbh}{IBBH}{intermediate binary black hole}
\newacronym{\bdj}{BDJ}{Brans-Dicke-Jordan}
\newacronym{\bbo}{BBO}{Big Bang Observer}
\newacronym{\decigo}{DECIGO}{Deci-Hertz Gravitational-Wave Observatory}

\newacronym{\git}{GT}{Georgia Institute of Technology}

\newacronym{\BSSNOK}{BSSNOK}{Baumgarte-Shapiro-Nakamura-Oohara-Kojima}
\newacronym{\adm}{ADM}{Arnowitt-Deser-Misner}

\def\MPR#1{{\it Moving Puncture Recipe}#1 (MPR#1)\gdef\MPR{MPR}}
\def\ahz#1{apparent horizon#1 (AH#1)\gdef\ahz{AH}}
\def\CM#1{center-of-mass#1 (CM#1)\gdef\CM{CM}}
\def\CLA#1{close-limit approximation#1 (CLA#1)\gdef\CLA{CLA}}
\def\pnw#1{post-Newtonian#1 (PN#1)\gdef\pnw{PN}}
\def\nr#1{numerical relativity#1 (NR#1)\gdef\nr{NR}}
\def\qnm#1{quasi-normal mode#1 (QNM#1)\gdef\qnm{QNM}}
\def\isco#1{innermost stable circular orbit#1 (ISCO#1)\gdef\isco{ISCO}}
\def\eos#1{equation of state#1 (EOS#1)\gdef\eos{EOS}}
\def\tov#1{Tolman-Oppenheimer-Volkoff#1 (TOV#1)\gdef\tov{TOV}}
\def\ns#1{neutron star#1 (NS#1)\gdef\ns{NS}}
\def\bbh#1{binary black hole#1 (BBH#1)\gdef\bbh{BBH}}
\newacronym{\wdbh}{WDBH}{white dwarf -- black hole}
\newacronym{\whd}{WD}{white dwarf}
\newacronym{\nswd}{NSWD}{neutron star -- white dwarf}
\newacronym{\gw}{GW}{gravitational wave}
\newacronym{\EM}{EM}{electromagnetic}
\newacronym{\srhd}{RHD}{relativstic hydrodynamics}
\newacronym{\grhd}{GRHD}{general relativstic hydrodynamics}
\newacronym{\gr}{GRHD}{General Relativity}
\newacronym{\pn}{PN}{post Newtonian}
\newacronym{\ehz}{EH}{event horizon}
\def\bhns#1{black hole -- neutron star#1 (BHNS#1)\gdef\bhns{BHNS}}
\def\nsns#1{neutron star -- neutron star#1 (NSNS#1)\gdef\nsns{NSNS}}
\def\emri#1{extreme mass-ratio inspiral#1 (EMRI#1)\gdef\emri{EMRI}}
\def\emrb#1{extreme mass-ratio binaries#1 (EMRB#1)\gdef\emrb{EMRB}} 
\def\grb#1{gamma-ray burst#1 (GRB#1)\gdef\grb{GRB}}
\def\imbh#1{intermediate mass black hole#1 (IMBH#1)\gdef\imbh{IMBH}}
\def\smbh#1{supermassive black hole#1 (SMBH#1)\gdef\smbh{SMBH}}
\def\bh#1{black hole#1 (BH#1)\gdef\bh{BH}}
\def\ulx#1{ultra-luminous x-ray source#1 (ULX#1)\gdef\ulx{ULX}}
\def\nps#1{Newman-Penrose#1 (NP#1)\gdef\nps{NP}} 
\def\lmxbs{low-mass x-ray Binaries (LMXBs)\gdef\lmxbs{LMXBs}\gdef\lmxb{LMXB}} 
\def\lmxb{low-mass x-ray Binary (LMXB)\gdef\lmxbs{LMXBs}\gdef\lmxb{LMXB}}

\newcommand{\DefineRun}[2]{%
  \expandafter\newcommand\csname run-#1\endcsname{#2}%
}
\newcommand{\Run}[1]{\csname run-#1\endcsname}

\DefineRun{b6a00}{B6S0} 
\DefineRun{b6ap6}{B6Su} 
\DefineRun{b6an6}{B6Sd} 
\DefineRun{b6a6t90}{B6Si} 
\DefineRun{b6a6t63p90}{B6Sa} 
\DefineRun{b8a6t63p90}{B8Sa} 

\shorttitle{Tidal disruption of white dwarfs by intermediate mass black holes}
\shortauthors{Haas, Shcherbakov, Bode, \& Laguna}

\begin{document}
\title{Tidal disruptions of white dwarfs from ultra-close encounters\\
  with intermediate mass spinning black holes}

\author{Roland Haas\altaffilmark{1,2}, Roman V. Shcherbakov\altaffilmark{3}, Tanja Bode\altaffilmark{1}, Pablo
  Laguna\altaffilmark{1}}

\altaffiltext{1}{Center for Relativistic Astrophysics, School of
  Physics, Georgia Institute of Technology, Atlanta, GA 30332, USA}
\altaffiltext{2}{Theoretical AstroPhysics Including Relativity,
California Institute of Technology, Pasadena, CA 91125, USA}
\altaffiltext{3}{Hubble Fellow, Department of Astronomy, University of
  Maryland, College Park, MD 20742-2421, USA}

\begin{abstract}
  We present numerical relativity results of tidal disruptions of
  white dwarfs from ultra-close encounters with a spinning,
  intermediate mass black hole.  These encounters require a full
  general relativistic treatment of gravity. We show that the
  disruption process and prompt accretion of the debris strongly
  depend on the magnitude and orientation of the black hole
  spin. However, the late-time accretion onto the black hole follows
  the same decay, $\dot M \propto\,t^{-5/3}$, estimated from Newtonian
  gravity disruption studies. We compute the spectrum of the disk
  formed from the fallback material using a slim disk model. The disk
  spectrum peaks in the soft X-rays and sustains Eddington luminosity
  for $1-3$~yrs after the disruption. For arbitrary black hole spin orientations, the disrupted material is
  scattered away from the orbital plane by relativistic frame
  dragging, which often leads to obscuration of the inner fallback
  disk by the outflowing debris. The disruption events also yield
  bursts of gravitational radiation with characteristic frequencies of
  $\sim 3.2$~Hz and strain amplitudes of $\sim 10^{-18}$ for galactic
  intermediate mass black holes. The optimistic rate of considered ultra-close 
  disruptions is consistent with no sources found in ROSAT all-sky survey. 
  The future missions like Wide-Field X-ray Telescope (WFXT) could observe 
  dozens of events.
\end{abstract}


\section{Introduction}\label{sec:introduction}
Tidal disruptions of stars by \bh{s} are fascinating and violent
cosmic events, releasing copious amounts of energy in electromagnetic
radiation and, in some cases, also accompanied by potentially
detectable gravitational wave emission. The scattered debris from a
stellar disruption will yield different radiation patterns depending
on the orientation of the orbit. This variety of radiation patterns
will provide clues about both the \bh{} and the internal structure of
the disrupted star.  As the debris from a tidal disruption showers
back into the \bh{} to form an accretion disk, for instance, it will
radiate close to or above the Eddington luminosity for some duration,
with the spectrum peaking at UV/X-ray frequencies depending on the
\bh{}
mass~\citep{Rees:1988bf,Evans:1989qe,Cannizzo1990,Bogdanovic:2004ac,Strubbe2009}. In
addition, the unbound debris will quickly scatter over a large volume,
and may be illuminated by the accretion disk, leading to an optical
irradiation spectrum dominated by
lines~\citep{Bogdanovic:2004ac,Sesana:2008zc,Strubbe2009,Clausen2011}.

When a main sequence star is disrupted, the super-Eddington outflow is
by itself hot enough to be a significant source of optical emission
resembling supernova radiation~\citep{Kasen2009,Strubbe2009}. In these
circumstances, instead of outflowing debris, a steady optically thick
envelope may reprocess the inner disk radiation
\citep{Loeb1997,Ulmer1998}. As a consequence, neither the inner disk
nor the irradiated debris emission is visible.  Another tidal
disruption effect arises from the tidal compression perpendicular to
the orbital plane.  The compression leads to the formation of a strong
shock that triggers a powerful short outburst
\citep{Kobayashi:2004py,Guillochon2009,Rosswog:2008ie}. In some
instances, e.g. ultra-close disruptions, the tidal compression could
be strong enough to produce a thermonuclear ignition of the star. In
such cases, the detonation of the star would be observed as an
under-luminous supernova \citep{Rosswog:2008ie}.

The variety of proposed radiative signatures often disagree with each
other, calling for precise dynamical modeling of tidal disruptions.
Early hydrodynamic simulations by \citet{Evans:1989qe} confirmed a
flat distribution of debris mass over its energy range, yielding the
predicted $t^{-5/3}$ law of fallback accretion
rate~\citep{Rees:1988bf}. Other numerical investigations addressed the
tidal compression and related shock formation; some of these studies
used smoothed particle hydrodynamics (SPH)
techniques~\citep{Kobayashi:2004py,Rosswog:2008ie} and others
hydrodynamic grid-based codes~\citep{Guillochon2009}. The studies by
\citet{Rosswog:2008ie} included nuclear reactions with simplified
networks. On the other hand, \citet{Bogdanovic:2004ac} looked at the
formation of spherical envelopes~\cite{Loeb1997} using SPH
simulations, but the study did not find any conclusive evidence of a
spherical structure. At present tidal disruption simulations that
capture all the relevant physics are quite challenging. The
simulations require handling, in addition to hydrodynamics, effects
from general relativity, nuclear reactions, and magnetic fields. Thus,
many important questions are still left unanswered. In particular,
there are very few simulations that account for general relativistic
effects
\citep{Laguna1993,Laguna1993a,Kobayashi:2004py,Bogdanovic:2004ac}.
The inclusion of magnetic fields may lead to jet formation.

Very likely, the \bh{s} involved in a tidal disruption will have spins
misaligned with the orbital angular momentum of the incoming
star. Thus the disk formed from the tidal debris will have angular
momentum that is also misaligned with the \bh{} spin
direction~\citep{Rees:1988bf}. For thin disks, the Bardeen-Petterson
effect will align the BH spin with the angular momentum of the inner
region of the disk \citep{Bardeen:1975zz}, but for the thick disks
expected from tidal disruptions, the alignment may not happen
\citep{Papaloizou:1983ge,Fragile:2007ac,Stone:2011hy}. This
misalignment could have important consequences on the emission
channels from the disruption, e.g. jet formation. The spin of the
\bh{} will also produce distinct effects in an ultra-close encounter,
when the pericentric radius $R_{\rm p}$ is comparable to the
gravitational radius $R_{\rm g}= G\, M_{\rm bh}/c^2$ of the \bh{.}  In
this strong gravity regime, the details of the disruption will depend
not only on the stellar radius, stellar equation of state, and orbital
energy, but also on the magnitude and orientation of the spin of the
\bh{.}  In particular, the properties of the accretion will depend on
the \bh{} spin since its magnitude determines the location of the
\isco{} radius $r_{\rm ISCO}$, ranging from $r_{\rm ISCO} = 9\,R_{\rm
  g}$ for a counter-rotating maximally spinning \bh{} to $r_{\rm ISCO}
=1R_{\rm g}$ for a co-rotating maximally spinning \bh{.}  The value of
spin determines whether the star following the same orbit gets
disrupted \citep{Kesden:2011a}.  Furthermore, a misaligned rotating
\bh{} will drag around nearby debris \citep{Bardeen1972,Shapiro1986},
and will effectively push this material away from the orbital
disruption plane. The net effect will be the formation of a shell-like
disruption debris engulfing the \bh{,} as opposed to the traditional
S-shape debris observed in Newtonian gravity
simulations~\citep{Evans:1989qe}.

The present work aims at exploring the exciting regime of ultra-close
encounters (i.e.  $R_{\rm p} \sim$ few $R_{\rm g}$) of a \whd{} with
an \imbh{}. Our main goal is to investigate observational signatures
due to strong gravity effects that may shed light on the presence of
\imbh{s}.  We consider a carbon-oxygen \whd{} with mass $M_{\rm
  wd}=1\,M_\odot$ and radius~\citep{1972ApJ...175..417N}
\begin{eqnarray}
  R_{\rm wd} &=& 1.56\,R_\oplus \left(\frac{M_{\rm wd}}{M_{\rm ch}}\right)^{-1/3}
\left[1-\left(\frac{M_{\rm wd}}{M_{\rm ch}}\right)^{4/3}\right]^{3/4}\nonumber\\
&=& 0.86\,R_\oplus
\end{eqnarray}
with $M_{\rm ch} = 1.44\,M_\odot$ the Chandrasekhar mass and $R_\oplus
= 6,960$ km the Earth mean radius.  We fix the mass of the \imbh{} to
$M_{\rm bh}=10^3\,M_\odot$. With these choices, $R_{\rm wd} \simeq
4\,R_{\rm g} \simeq 6,000\,\mathrm{km}$. Therefore, the \whd{} and
\imbh{} can be numerically modeled with comparable grid
resolutions. The other important scale in the problem is the tidal
disruption radius $R_{t}$. For our setup \citep{Rees:1988bf},
\begin{equation}
\label{eqn:tidalradius}
\frac{R_{\rm t}}{R_{\rm g}} \simeq 40\,\left(\frac{R_{\rm wd}}{0.86\,R_\oplus}\right)
\left(\frac{M_{\rm wd}}{1\,M_\odot}\right)^{-1/3}
\left(\frac{M_{\rm bh}}{10^3\,M_\odot}\right)^{-2/3} \,,
\end{equation}
gave us ample room to carry out deep penetration encounters.

Currently, there is a large body of evidence for the existence of both
solar mass as well as supermassive \bh{s} with masses in the range of
$10^5$ -- $10^8\,\msun$. \imbh{s} are, however, the missing link
between stellar mass and supermassive \bh{s}. Today, only tentative
evidence exists~\citep{Irwin:2009jc,Farrell:2010bf,Davis:2011ka}. On
the other hand, \whd{s} are thought to be abundant in spiral
galaxies~\citep{Evans:1987qa,Reid:2005} and globular
clusters~\citep{Gerssen:2002iq}. Thus the identification of distinct
signals in both \gw{s} and the electromagnetic
spectrum~\citep{Gould:2010ij} from a disruption event could
potentially guide observations that provide evidence for the existence
of \imbh{} as well as insights into the structure of the \whd{}
involved.

Our study uses the full machinery of numerical relativity to solve the
Einstein equations of general relativity and hydrodynamics. We do not
include nuclear reactions. For the ultra-close encounters of our
interest, with $R_{\rm p} \sim$ few $R_{\rm g}$, a general
relativistic description of gravity is needed. However, given the mass
ratio $M_{\rm wd}/M_{\rm bh} \simeq 10^{-3}$ of the bodies involved
and encounters not driven by \gw{} emission, one does not need to
account for dynamical general relativistic gravity.  The present study
could have been carried out by doing hydrodynamics on the fixed
space-time background provided by the \imbh{.} However, this would
have implied developing a general relativistic hydro code on a fixed
background, a code similar to the SPH code developed by one of us to
investigate the tidal disruption of main sequence stars by
supermassive \bh{s}~\citep{Laguna1993,Laguna1993a}.  We decided
instead to take advantage of our numerical relativity code \maya{.}
The code has demonstrated excellent performance in handling fluid
flows in the vicinity of \bh{s} in our studies of binary \bh{} mergers
in astrophysical environments~\citep{Bode2008,Bode2009a,Bode:2011ac}.
The other advantage of using the \maya{} code is the ability to obtain
the \gw{} signal directly from the simulation, without having to take
recourse to the possibly inaccurate quadrupole formula or the
application of \pn{} approximations to space-times containing a
spinning \bh{.}

The paper is organized as follows.  In \S~\ref{sec:initial_data}, we
describe how self-consistent general relativistic initial data is
constructed.  Results from the dynamics of the disruption events are
discussed in \S~\ref{sec:results}. We estimate electromagnetic
transient radiation in \S~\ref{sec:radiation} using a slim disk model
to compute the spectrum during the fallback phase. We find that the
sources shine at Eddington luminosity $\sim10^{41}{\rm erg~s}^{-1}$
for about $1-3$~yrs, and then fade approximately as $\propto
t^{-5/3}$.  We also find that there is a 50/50 chance that the inner
disk will be obscured by outflowing debris for a fully misaligned
\bh{} spin.  In \S~\ref{sec:gravitational-waves}, we calculate the
gravitational wave signal produced during the encounters.  
In \S~\ref{sec:observations} we estimate the event rates and 
discuss the observational prospects. We discuss the progress and the limitations of our study in
\S~\ref{sec:conclusions}.

\section{Initial data and code tests}\label{sec:initial_data}

When using a fully general relativistic code, such as our \maya{}
code, that is based on a 3+1 formulation of the Einstein equations,
the initial data for the dynamical space-time are comprised of the
spatial metric $\gamma_{ij}$ and the extrinsic curvature $K_{ij}$ of
the initial space-time hypersurface. All tensor indices are spatial
indices and, in this section only, we use units for which $G=c=1$. The
metric $\gamma_{ij}$ characterizes the gravitational potentials of the
\imbh{} and \whd{,} and the tensor $K_{ij}$ provides the ``velocity''
of the metric, or the embedding of the space-like hypersurface, in the
space-time.\footnote{For a review on numerical relativity, we
  recommend the textbooks by \citet{Alcubierre:2008} and
  \citet{Baumgarte:2010ab}}

Although $\gamma_{ij}$ and $K_{ij}$ are dominated by the \imbh{,} it
is crucial to account for the contributions from the \whd{} in order
to correctly include the self-gravity of the \whd{.} The components of
$\gamma_{ij}$ and $K_{ij}$ cannot all be freely specifiable. They need
to satisfy the following elliptic equations
\begin{align}
    8\, \Laplacian \phi + \phi^{-5} \bar A_{ij} \bar A^{ij} &= -16\,\pi\, \phi^{-3}\bar \rho_*
    \label{eqn:hamiltonian-constraint}\text{,} \\
   \partial_j \bar A^{ij} &= 8\,\pi \,\bar J^i\label{eqn:momentum-constraint} \text{,}
\end{align}
where
\begin{eqnarray}
\gamma_{ij} &=& \phi^4 \delta_{ij}\\
K^{ij} &=& \bar A^{ij} \phi^{-10}\\
\rho_* & = &   \bar \rho_* \phi^{-8} \label{eqn:conformal-rho} \\
J^i &=&    \bar J^i \phi^{-10} \label{eqn:conformal-S}\,.
\end{eqnarray}
Above, $\phi$ is a conformal factor, $\delta_{ij} =
\text{diag}(1,1,1)$ the flat spatial metric and $ \Laplacian$ its
associated Laplace operator. In writing
Eqs.~(\ref{eqn:hamiltonian-constraint}) and
(\ref{eqn:momentum-constraint}), we have made the customary assumption
that the initial physical metric $\gamma_{ij}$ is conformally flat
and $K_{ij}$ is trace-free.  Eqs.~(\ref{eqn:hamiltonian-constraint})
and (\ref{eqn:momentum-constraint}) are, respectively, the conformal
versions of the so-called Hamiltonian and momentum constraints of
general relativity~\citep{York:1979}.  The sources $\rho_*$ and $J^i$
are, respectively, the total energy and momentum densities.

Our approach to solve the Hamiltonian and momentum constraints is to
specify the ``free'' data in Eqs.~(\ref{eqn:hamiltonian-constraint})
and (\ref{eqn:momentum-constraint}) from the solutions to an isolated,
spinning \bh{,} and a boosted \whd{} \citep{Loffler:2006wa}.

Let us consider first the solution for a boosted \whd{.}
The sources $\bar
\rho_*$ and $\bar J^i$ in Eqs.~(\ref{eqn:hamiltonian-constraint}) and
(\ref{eqn:momentum-constraint}) are obtained from the stress-energy
tensor for a perfect fluid (see chapter 5 in
\citealt{Baumgarte:2010ab}). They read
\begin{align}
    \bar\rho_* &= \rho \,\left(1 + \varepsilon + \frac{P}{\rho}\right) W^2 - P
    \text{,}\label{eqn:conserved-energy}\\
    \bar J^i    &= \rho\, \left(1 + \varepsilon + \frac{P}{\rho}\right) W^2 v^i
    \text{,}\label{eqn:conserved-current}
\end{align}
where $W = (1-\gamma_{ij}v^i v^j)^{-1/2}$ is the Lorentz factor and
$v^i$ the boost velocity of the \whd{.}  In
(\ref{eqn:conserved-energy}) and (\ref{eqn:conserved-current}), $\rho$
is the rest-mass density, $P$ the pressure, and $\varepsilon$ the
specific internal energy density.  The values for those quantities are
obtained from \tov{}
solutions~\citep{Oppenheimer:1939ne,Tolman:1939jz} with a polytropic
equation of state $P = K \, \rho^\Gamma$.
We use $\Gamma = 5/3$ appropriate for a non-degenerate gas, which is
sufficient to capture the dynamics during the inspiral and disruption
phases.
During the evolution, we
switch to a gamma-law equation of state $P = \rho \, \varepsilon \,
(\Gamma-1)$. We denote the solutions to
Eqs.~(\ref{eqn:hamiltonian-constraint}) and
(\ref{eqn:momentum-constraint}) in the absence of the \bh{} by
$\phi_{\rm wd}$ and $\bar A^{\rm wd}_{ij}$.

Now we consider the solution to the constraints for a spinning~\bh{.}
In the absence of the \whd{,} $\bar\rho_* = 0$ and $\bar J^i = 0$. In
this case, the momentum constraint (\ref{eqn:momentum-constraint}) can
be solved analytically~\citep{1980PhRvD..21.2047B}. The solution reads
\begin{equation}
\label{eqn:bowen}
   \bar A^{\rm bh}_{ij} = \frac{3}{r^3}\left(\epsilon_{kil} S^l l^k l_j + \epsilon_{kjl}
                   S^l l^k l_i\right)
\end{equation}
with $l^i = x^i/r$ and $\epsilon_{ijk}$ the three-dimensional
Levi-Civita symbol. Above, $S^i$ is the spin vector of the \bh{.}

With the solutions $\bar A^{\rm wd}_{ij}$ and $\bar A^{\rm bh}_{ij}$
at hand, it is clear that the linear superposition $\bar A_{ij} = \bar
A^{\rm bh}_{ij} + \bar A^{\rm wd}_{ij}$ is the solution to the
momentum constraint (\ref{eqn:momentum-constraint}) in the presence of
both a \whd{} and a \bh{.} Next, we solve the Hamiltonian constraint
(\ref{eqn:momentum-constraint}) with $\bar A_{ij} = \bar A^{\rm
  bh}_{ij} + \bar A^{\rm wd}_{ij}$ and $\bar\rho_*$ given by
(\ref{eqn:conserved-energy}). We use the popular ansatz
\begin{equation}
   \label{eqn:full}
   \phi = 1+\frac{M_{\rm bh}}{2\,r} +u\,.
\end{equation}
Thus, equation~(\ref{eqn:hamiltonian-constraint}) becomes an equation
for $u$.  In (\ref{eqn:full}), the first term is the flat space
solution to (\ref{eqn:hamiltonian-constraint}) ; the second is the
\bh{} solution; and the third term $u$ includes both the gravitational
potential of the \whd{} as well as the interaction effects with the
\bh{.}

\begin{figure}[htbp]
    \centering\plotone{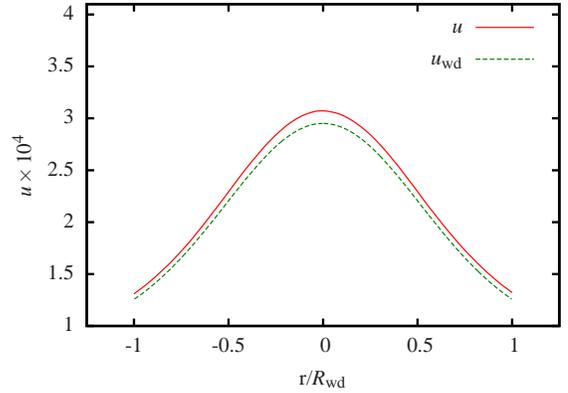}
    \caption{Solution $u$ to equation~(\ref{eqn:222}) for a \whd{} at rest
      a distance 15 $R_{\rm wd}$ away from a non-spinning \imbh{.}
      The solution $u$ provides not only the gravitational potential
      of the \whd{} but also includes the full non-linear interactions
      with the \imbh{.} For comparison, we plot also $u_{\rm wd}$, the
      solution to equation~(\ref{eqn:222}) for a \whd{} in isolation.}
    \label{fig:non-linear-superposition}
\end{figure}

To test our initial data method, and in particular to assess the
importance of the \whd{-}\bh{} interaction effects, we compute initial
data for a \whd{} initially at rest; that is, $\bar J^i = 0$ and thus
$\bar A^{\rm wd}_{ij}= 0$. In addition, we assume a non-spinning
\bh{,} i.e. $\bar A^{\rm bh}_{ij}= 0$. The Hamiltonian constraint
\eqref{eqn:hamiltonian-constraint}, with the help of
(\ref{eqn:full}), takes the form
\begin{align}
\label{eqn:222}
    \bar \Laplacian u &= -16\, \pi \,\phi^{-3}\bar \rho_* \,.
\end{align}
The solution to (\ref{eqn:222}) is shown in
Figure~\ref{fig:non-linear-superposition}. For color versions of the
figures in this paper see the electronic edition of the Journal. Also
in Figure~\ref{fig:non-linear-superposition} is the solution $u_{\rm
  wd}$ to equation~(\ref{eqn:222}) without the \bh{;} that is, with
$M_{\rm bh} = 0$ in (\ref{eqn:full}).  The difference between $u$ and
$u_{\rm wd}$ is entirely due to the gravitational interaction of the
\whd{} with the \bh{.}

We also tested the ability of our method to produce stable isolated
\whd{} models and to conserve rest mass.  We found that, over six
dynamical times, the central density of the \whd{} did not change more
than $1.2\%$ of its initial value and rest mass was conserved at the
$0.1\%$ level. This is in spite of the adaptive mesh refinement
infrastructure we use (\carpet{)} not being designed to accurately
preserve rest mass conservation when interpolating between different
refinement levels.

\section{Simulation Results}\label{sec:results}

\subsection{Tidal Disruption Simulation Parameters}

\begin{figure}[htbp]
  \centering\plotone{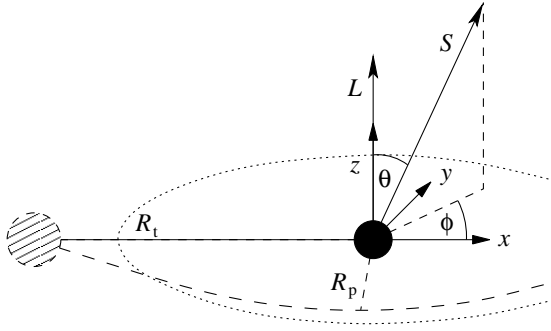}
  \caption{Configuration setup of our simulations. The \whd{} is
    located a distance $1.5 \,R_{\rm t} \simeq 60\,R_{\rm g}$ along
    the negative $\hat x$-axis away from the \bh{.} The orbital
    angular momentum $\vec L$ of the \whd{} is aligned with the $\hat
    z$-axis. The orientation of the \bh{} spin $\vec S$ is determined by the
    standard $\theta$ and $\phi$ angles in a right handed spherical
    coordinate system. }
    \label{fig:setup}
\end{figure}

Our study consists of a series of six simulations. As mentioned
before, the \imbh{} has a mass $M_{\rm bh} = 10^3\,M_\odot$, and the
\whd{} has a mass $M_{\text{wd}} = 1 \,M_\sun$ and radius
$R_{\text{wd}} \simeq 6,000\,\mathrm{km}$~\citep{Hamada:1961aa}. The
\whd{} mass and radius correspond to a central density
$\rho_{\text{initial}} = 1.33\times10^7\,\mathrm{g}\,\mathrm{cm}^{-3}$,
$\Gamma = 5/3$, and polytropic constant $K = 2.41\times10^{-9}
\,(\mathrm{cm}^3\,\mathrm{g}^{-1})^{\Gamma-1}$. All the simulations
start with the \whd{} in a parabolic orbit, in the Newtonian gravity
sense, a distance $1.5 \,R_{\rm t} \simeq 60\,R_{\rm g}$ along the
negative $\hat x$-axis away from the \imbh{.} The orbital angular
momentum of the \whd{} is aligned with the $\hat z$-axis, as depicted
in Figure~\ref{fig:setup}.

Table~\ref{tab:initial-configurations} lists the configurations used
in the simulations where $\beta = R_{\rm t}/R_{\rm p}$ is the penetration
factor for an orbit with pericentric distance $R_{\rm p}$, and $a^\star = |\vec
S|/M^2$ is the magnitude of the dimensionless spin parameter of the
\imbh{.} The angles $\theta$ and $\phi$ determine the direction of the
\bh{} spin. They are the standard angles in a right-handed spherical
coordinate system as depicted in Figure~\ref{fig:setup}. We use the
following convention to label the cases we studied.  If a case is
labeled "BxSy", the \whd{} had penetration factor $\beta=x$. In
addition, the value of y denotes a non-spinning \bh{} (y=0) or the
\bh{'s} spin orientation: (up, down, in-plane, arbitrary) for y = (u,d,i,a),
respectively.
\begin{table}[htbp]
    \centering%
    \begin{tabular}{lccccc}
        Run & $\beta$ & $a^\star$ & $\theta$ & $\phi$ \\
        \hline
        \Run{b6a00} & 6 & 0.0 & 0\cdeg & 0\cdeg\\
        \Run{b6ap6} & 6 & 0.6 & 0\cdeg & 0\cdeg\\
        \Run{b6an6} & 6 & 0.6 & 180\cdeg & 0\cdeg\\
        \Run{b6a6t90} & 6 & 0.6 & 90\cdeg & 0\cdeg\\
        \Run{b6a6t63p90} & 6 & 0.6 & 63\cdeg & 90\cdeg\\
        \Run{b8a6t63p90} & 8 & 0.6 & 63\cdeg & 90\cdeg
    \end{tabular}
    \caption{Simulation parameters: $\beta = R_{\rm t}/R_{\rm p}$ is the penetration factor,
      $a^\star = |\vec S|/M^2$ is the dimensionless spin parameter
      of the central black hole, $\theta$ and $\phi$ are the angles between
      $\vec S$ and the orbital angular momentum ($\hat z$-axis) and the radial
      direction joining the \imbh{} and the \whd{} ($\hat x$-axis),
      respectively.}
    \label{tab:initial-configurations}
\end{table}

In all the simulations, we employ eight levels of mesh refinement with
radii $1.24\,R_{\rm g} \times 2^\ell$ ($0 \le \ell \le 7$) centered at
the \bh{} location. In addition to these refinement levels, we
surround the \whd{} during the pre-disruption phase with five
additional nested boxes of radii $4.96\,R_{\rm g} \times 2^\ell$ ($0
\le \ell \le 4$).  The resolution on the finest refinement is $R_{\rm
  g}/19.35$, the resolution on the level covering the \whd{} is
$R_{\rm g}/9.675$. This mesh refinement setup becomes insufficient
when the \whd{} begins to be disrupted. At this point, we turn off the
boxes tracking the \whd{} and construct a larger mesh with resolution
$R_{\rm g}/9.675$ that includes both the \imbh{} and the highly
distorted \whd{.}  Once the expanding debris has cleared the inner
region and the density has dropped, we turn off this level again to
speed up the simulation.

\begin{figure}[htbp]
    \centering\plotone{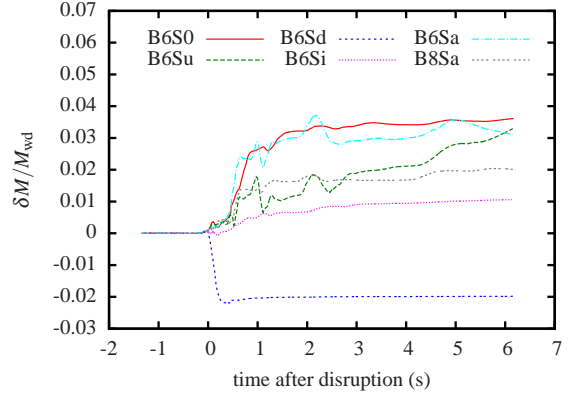}
    \caption{Relative gain/loss of rest mass in the computational
      domain, $\delta M/M_{\rm wd} = (M - M_{\rm wd})/M_{\rm wd}$, where $M$ is
      the total relativistic rest mass and $M_{\rm wd}$ the initial
      \whd{} rest mass.  The total relativistic rest mass $M$ takes
      into account fluxes into the \bh{.} The only simulation run
      displaying a net mass loss is the case \Run{b6an6} in which
      almost all the material accretes promptly. In the other cases,
      most of the mass gain happens in the short high compression
      phase when the \whd{} is at its periapsis point close to the
      \bh{.}}
    \label{fig:mass-conservation}
\end{figure}

The mesh refinements affect our ability to conserve rest mass
throughout the computational domain, in particular if the meshes are
moving, created, or removed. Figure~\ref{fig:mass-conservation}
displays the relative gain/loss of rest mass in the computational
domain, $\delta M/M_{\rm wd} = (M - M_{\rm wd})/M_{\rm wd}$, where $M$ is the
total relativistic rest mass and $M_{\rm wd}$ the initial \whd{} rest
mass.  The computed value of $M$ takes into account the mass loss
through the \bh{} horizon and the outer boundary of the computational
domain. We find tolerable mass conservation, with violations of less
than $4\%$ over the course of the simulations. Runs \Run{b6ap6} and
\Run{b6a00} display the worst mass conservation. As we shall see,
these are the cases in which most of the material of the star
($\gtrsim 95\%$) escapes the \bh{,} forming an expanding cloud that
reaches the coarsest mesh refinements.  Notice also that most of the
errors in mass conservation accumulate after the disruption.  Run
\Run{b6an6} is the only run that loses mass. This is the case in which
the \imbh{} accretes almost all of the star during the initial passage
of the \whd{.}

\subsection{Tidal compression}\label{sec:tidal-compression}
\begin{figure}[htbp]
    \centering\plotone{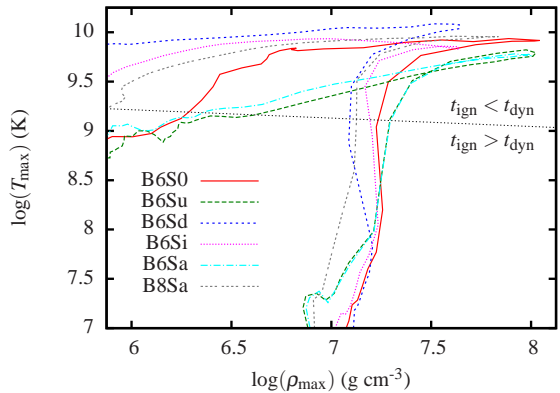}
    \caption{Instantaneous maximum temperature as a function of the
      density at the same location. We compute the density-weighted
      temperature average of the hottest cells in the grid which make
      up $10\%$ of the total rest mass of the \whd{.} The dotted black
      line separates regions in parameter space where the ignition time
      as estimated from Figure~2 of~\cite{Dursi:2005jg} is larger/smaller
      than the dynamical timescale of the \whd{,} i.e. the region of
      successful explosion from the region where the explosion fails.
      Densities smaller than $10^7\,\mathrm{g}\,\mathrm{cm}^{-3}$ are not
      present in Figure~2 and the estimates for the ignition time in this
      region are based on simple linear extrapolation.
      In all cases investigated, the ignition timescale
      $t_{\mathrm{ign}}$ of the \whd{} is
      much shorter than the dynamical timescale $t_{\mathrm{dyn}} =
      1/\sqrt{G\rho} \gtrsim 0.38\,\mathrm{s}$ of the \whd{.}}
    \label{fig:temp-vs-rho}
\end{figure}

The first stage in a tidal disruption encounter consists of the
deformations that the incoming star experiences due to the tidal
forces. In the orbital plane of the star, there are two competing
tidal deformations. One of them is tidal stretching along the radial
direction joining the \imbh{} with the \whd{.} The other is tidal
compression perpendicular to this radial direction.  In the direction
perpendicular to the orbital plane, there is also tidal
compression. This compression, or squeezing, is stronger than that in
the orbital plane because in the latter the radial direction changes
rapidly as the star reaches periapsis, thus modifying the ``contact
point'' in the star of the in-plane compression.  We concentrate here
on the compression in the direction perpendicular to the orbital
plane. This compression could, in principle, generate enough heat to
detonate the
star~\citep{1985AnPh...10..101L,1985MNRAS.212...57L,1989A&A...209...85L}
by igniting nuclear reactions at the core of the
\whd{}~\citep{Rosswog:2008ie}.  To investigate whether a nuclear
ignition is likely to happen from tidal compression, we plot in
Figure~\ref{fig:temp-vs-rho} the instantaneous maximum temperature of
the star as a function of the density at that location. To a good
approximation, this density is also the instantaneous maximum density.
The instantaneous maximum temperature is computed as the
mass-density-weighted temperature average of the portion of the star
whose numerical cells that have the largest internal energy and make
up for $10\%$ of its mass.

We calculate temperatures $T$ from the thermal specific energy as
(see~\cite{Lee:2005se})
\begin{align}
    m_p n_{\rm nuc} \varepsilon_{\text{th}} &=
      \frac32 k_B T (N_{\rm ion} + N_e) n_{\rm nuc} +
      a_{BB} T^4 \text{,}
\end{align}
with
\begin{align}
    \varepsilon_{\text{th}} &= \varepsilon - \frac{K_{\text{initial}}}{\Gamma-1} \rho^{\Gamma-1}
    \label{eqn:thermal-temperature}\text{,}
\end{align}
where the second term in equation~\eqref{eqn:thermal-temperature} is
the ``cold'' specific internal energy in the absence of shocks. $N_e$
is the number of electrons per nucleon making up the gas, $N_{\rm
  ion}$ is the number of ions per nucleon, $n_{\rm nuc}$ is the number
density of nucleons in the gas, $m_p$ is the proton (nucleon) mass,
$k_B$ is the Boltzmann constant, and $a_{BB}$ here is the radiation
constant. For simplicity we assume $N_{\rm ion} = 1/14$, for
an equal mixture of oxygen and carbon atoms.  For $N_e$, which depends
on the ionization state of the plasma, we use $N_e = 0$ for $T
\lesssim 5\times10^6\,\mathrm{K}$ and $N_e = 1/2$ for
$5\times10^6\,\mathrm{K} \lesssim T \lesssim
1\times10^{10}\,\mathrm{K}$.
That is, we ignore any partial ionization states of the constituent
atoms at low temperatures.

In all cases investigated, the maximum temperature and density
displayed in the tracks of Figure~\ref{fig:temp-vs-rho} reach high enough
values to trigger nuclear burning in the star~(cf.\ Figure 2 of
\cite{Dursi:2005jg}). Interestingly, the temperature-density tracks in
Figure~\ref{fig:temp-vs-rho} are qualitatively similar to those
by~\citet{Rosswog:2008ie}, who included nuclear networks.

\begin{table}[htbp]
    \centering%
    \begin{tabular}{lccccc}
        Run
        & $\beta^*$
        & $T_{\rm max}$ ($10^9\mathrm{K}$)
        & $\rho_{\rm max}/\rho_{\text{initial}}$
        & $f_{\text{acc}}$
        & $f_{\text{unb}}$ \\
        \hline
        \Run{b6a00}      & $8.9$ & $8.6$ & $7.5$ & $68\mathrm{\%}$ & $19\mathrm{\%}$ \\
        \Run{b6ap6}      & $9.44$ & $6.6$ & $9.3$ & $<1\mathrm{\%}$ & $60\mathrm{\%}$  \\
        \Run{b6an6}      & $11$ & $12 $ & $3.2$ & $>99\mathrm{\%}$& $<0.5\mathrm{\%}$  \\
        \Run{b6a6t90}    & $7.6$ & $8.6$ & $3.3$ & $65\mathrm{\%}$ & $22\mathrm{\%}$  \\
        \Run{b6a6t63p90} & $9.1$ & $6.0$ & $8.1$ & $2\mathrm{\%}$  & $67\mathrm{\%}$  \\
        \Run{b8a6t63p90} & $10$ & $9.1$ & $4.0$ & $43\mathrm{\%}$ & $34\mathrm{\%}$  
    \end{tabular}
    \caption{Actual penetration factor $\beta^*$ measured from the simulations. Maximum temperature $T_{\rm max}$ and compression
      $\rho/\rho_{\text{initial}}$ with $\rho_{\text{initial}} =
      1.33\times10^7\,\mathrm{g}\,\mathrm{cm}^{-3}$. Fraction $f_{\text{acc}}$ of the star accreted during the first $2\mathrm{s}$
      after disruption, unbound fraction $f_{\text{unb}}$ at the end of the simulation at
      $\sim6\mathrm{s}$.
    }
    \label{tab:accreted-fraction}
\end{table}

\begin{figure}[htbp]
    \centering\plotone{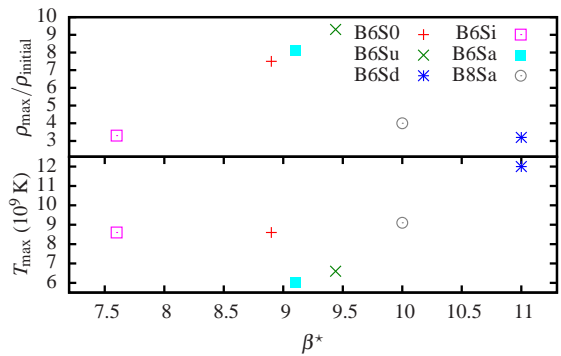}
    \caption{Maximum density $\rho_{\rm max}$ (top) and temperature $T_{\rm
    max}$ as a function of $\beta^*$ for the values in
    Table~\ref{tab:accreted-fraction}.}
    \label{fig:penetration}
\end{figure}

In Table~\ref{tab:accreted-fraction}, we list the maximum temperature
$T_{\text{max}}$ and compression
$\rho_{\text{max}}/\rho_{\text{initial}}$. Here $T_{\text{max}}$ and
$\rho_{\text{max}}$ are the maximum temperature and density attained
over the course of the evolution. We also list the ``actual''
penetration factor $\beta^* = R_{\rm t}/R_{\rm p}^*$.  Recall that the
orbital parameters, and in particular $R_{\rm p}$, used to set the
\whd{} in a parabolic orbit were obtained using Newtonian gravity.
Because of relativistic effects, the value of $R_{\rm p}^*$ will
differ from the Newtonian estimate $R_{\rm p}$, and thus $\beta \ne
\beta^*$.  In our simulations, we define $R_{\rm p}^*$ as the distance
of closest approach of the point within the \whd{} where $T_{\rm
  max}$, or equivalently $\rho_{\rm max}$, is found.  Notice that for
the \Run{b6ap6} and \Run{b6an6} cases, $T_{\rm max}$ almost doubles
and $\rho_{\rm max}$ triples when going from a spin aligned with the
orbital angular momentum to one that is anti-aligned. This is because
in the anti-aligned case (\Run{b6an6}), the \whd{} penetrates closer,
$\beta^* = 11$, than in the aligned case (\Run{b6an6}), $\beta^* =
9.4$.  Figure~\ref{fig:penetration} shows the maximum temperature
$T_{\rm max}$ and maximum density $\rho_{\rm max}$ as a function of
the actual penetration factor $\beta^*$ from
Table~\ref{tab:accreted-fraction}. Because of the small range of
values covered by $\beta^*$, it is not possible to verify the scaling
$T_{\rm max} \propto \beta^2$ and $\rho_{\rm max} \propto \beta^3$
suggested by~\citet{1982Natur.296..211C}

\subsection{Outflowing debris and spin}
\label{sec:shock-formation-and-heating-of-debris}

All cases showed that the mass and orientation of the debris after
disruption is strongly affected by the \bh{} spin. This is because the
spin determines the location of \isco{} radius, and it is also
responsible for frame-dragging effects.  In particular, frame-dragging
pulls the material out of the orbital plane, thereby changing the
orientation of the resulting disk.

We first discuss three disruptions: disruption by a non-spinning
\imbh{} (case \Run{b6a00}), by an \imbh{} with spin aligned (case
\Run{b6ap6}), and by an \imbh{} with spin anti-aligned (case
\Run{b6an6}) relative to the orbital angular momentum (see
Table~\ref{tab:accreted-fraction}).  For the spinning \bh{} cases, the
magnitude of the dimensionless spin parameter $a^*$ was kept at $0.6$.
With this spin magnitude, $r_{\rm ISCO} = 5,670\,\mathrm{km}$ for the
case \Run{b6ap6}, and $r_{\rm ISCO} = 11,600\,\mathrm{km}$ for the
case \Run{b6an6}. In the non-spinning case, $r_{\rm ISCO} =
9,000\,\mathrm{km}$~\citep{2011CQGra..28k4009M}.  All three cases had
penetration factors $\beta=R_{\rm t}/R_{\rm p}=6$.  From
equation~(\ref{eqn:tidalradius}), $R_{\rm t} = 40\,R_{\rm g} =
60,000\,\mathrm{km}$, thus a penetration factor of 6 translates to a
pericentric distance $R_{\rm p} \simeq 10,000\,\mathrm{km}$.  We then
expect that the \whd{} in the retrograde case \Run{b6an6} will pass
within the \isco{;} in the prograde case \Run{b6ap6} the star will
mostly stay outside of the \isco{;} and in the case \Run{b6a00} the
\whd{} will graze by the \isco{.}  Because $r_{\rm ISCO} > R_{\rm p}$
in the \Run{b6an6} case, almost all of the star is accreted by the
\bh{} soon after the disruption. For the cases \Run{b6a00} and
\Run{b6ap6}, on the other hand, most of the stellar debris escapes
direct capture, expanding away from the hole. The debris cloud has a
shape resembling a thick circular arch.  This arch-like cloud
eventually closes up and forms a disk around the \bh{.}

\begin{figure*}[htbp]
    \centering\plotone{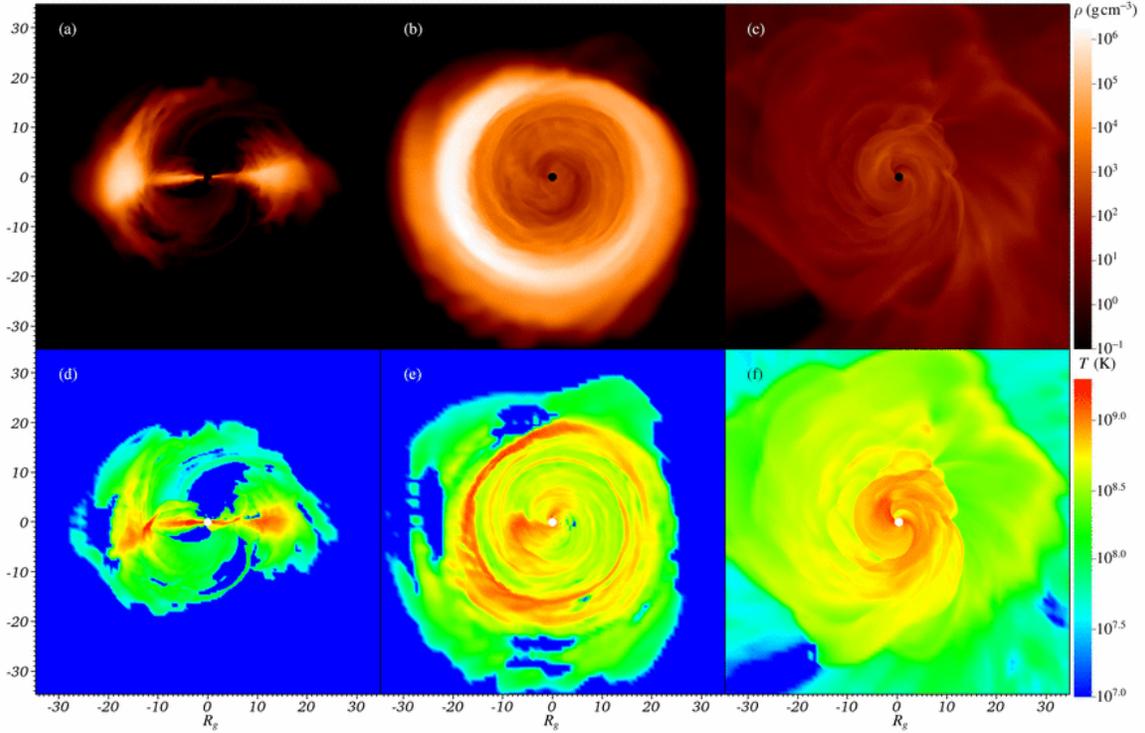}
    \caption{Snapshots in the orbital plane of the
density (top row, in $\mathrm{g}\,\mathrm{cm}^{-3}$) and temperature (bottom row,  in $\mathrm{K}$) for runs \Run{b6a6t90} and \Run{b6ap6}.
The snapshots in the left column are for the case \Run{b6a6t90} $\sim 0.5$ seconds after disruption.
The middle and right columns correspond to case \Run{b6ap6} $\sim 0.5$ seconds  and
$\sim 5.9$ seconds after disruption, respectively. }
    \label{fig:2D-temperature-rho}
\end{figure*}

In all cases, the leading edge of the tidally-disrupted star wraps
around the \bh{} and crashes into the
trailing material creating a hot region. For the \Run{b6ap6}, \Run{b6an6} and \Run{b6a00} cases,
this hot region is mostly contained within the equatorial plane and
moves outwards from the central region in the form of a spiraling
disturbance in the debris disk, eventually appearing on the debris surface.
 Figure~\ref{fig:2D-temperature-rho} shows snapshots in the orbital plane of the
density (top row) and temperature (bottom row) for runs \Run{b6a6t90} and \Run{b6ap6}.
The snapshots in the left column are for the case \Run{b6a6t90} $\sim 0.5$ seconds after disruption.
The middle and right columns correspond to case \Run{b6ap6} $\sim 0.5$ seconds  and
$\sim 5.9$ seconds after disruption, respectively.    In the aligned case \Run{b6ap6}, the hot region forms a spiral
      which moves outwards from the point where the leading edge of
      the \whd{} intersected with the tail material. At late times this
      feature has been washed out. In run \Run{b6a6t90} the hot region
      is much smaller; frame dragging by the \bh{} has pulled material
      out of the plane into a shell surrounding the \bh{.}

For the cases in which the spin of the \bh{} is not aligned with
the orbital angular momentum of the binary (i.e. \Run{b6a6t90}, \Run{b6a6t63p90} and
\Run{b8a6t63p90} cases), frame dragging is effectively able to wrap
the debris material around the \bh{.} This effect -- which is not
captured by Newtonian calculations -- greatly changes the geometry of
the debris, blanketing the central region with a shell-like cloud of
gas. We will address the observational consequences of the material
surrounding the \bh{} in the next section. The effect of the misalignment is evident from Figure~\ref{fig:2D-temperature-rho} in the remarkable differences of the spiral patterns
observed between the left and middle snapshots at $\sim 0.5$ seconds after disruption.
The left panels are for the \Run{b6a6t90} case in which the \bh{} spin direction is perpendicular to the orbital angular momentum. On the other hand, the middle panels are for the
\Run{b6ap6} case in which the \bh{} spin is aligned with orbital angular momentum. The observed differences are a consequence of  how the debris responds to the orientation of the
\bh{} spin.

\subsection{Prompt and late accretion}\label{sec:prompt-accretion}

We observe an initial prompt phase of accretion when the star swings
by the \bh{.} This prompt accretion phase is followed by an
intermediate phase in which material initially deflected by the \bh{}
slows down and begins to accrete. At late times, long after the end of our
simulations, the characteristic $t^{-5/3}$ fallback behavior takes
over.

\begin{figure}[htbp]
    \centering\plotone{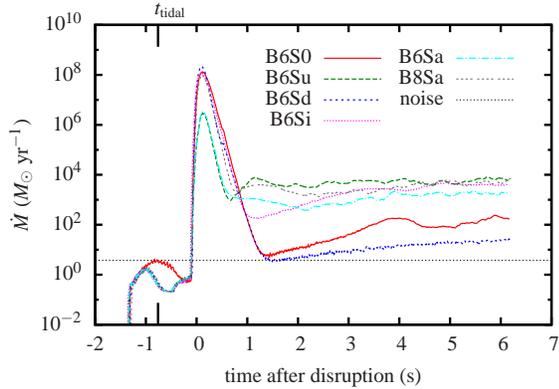}
    \caption{Accretion rate during the simulation. $t_{\text{tidal}}$
      indicates the approximate time when the \whd{} enters within the
      tidal radius $R_{\rm t}$.  The horizontal dotted black line indicates the mass
      accretion which we observe initially around the black hole. It
      is a purely numerical artifact from the atmosphere used to model
      the vacuum regions.}
    \label{fig:prompt-accretion}
\end{figure}
Figure~\ref{fig:prompt-accretion} shows the flow of matter through the
\imbh{'s} horizon during the first six seconds after the disruption
takes place.  The instant labeled $t_{\text{tidal}}$ is the time at
which the star enters the tidal radius. The horizontal dashed line
labeled ``noise'' denotes the level of accretion due to the atmosphere
used to model the vacuum regions. We observe that the magnitude of the
prompt accretion rate decreases the more prograde an orbit is. Both
\Run{b6a00} and \Run{b6ap6} disrupt at comparable distances from the
central \bh{,} but the peak accretion rate differs by almost two orders of magnitude.

As already pointed out in
Section~\ref{sec:shock-formation-and-heating-of-debris}, the \bh{} spin
strongly affects what fraction of the star is accreted onto the \bh{}
soon ($\sim 2$ s) after the star
disrupts. Table~\ref{tab:accreted-fraction} shows this accreted
fraction for each case. In the \Run{b6an6} case, the star is
completely accreted. On the other hand, the debris in the \Run{b6ap6}
and \Run{b6a6t63p90} cases escapes almost in its entirety during the
flyby. Notice also from Table~\ref{tab:accreted-fraction} that the
prompt accretion is also correlated with the maximum density
$\rho_{\rm max}$ the \whd{} is able to reach as a consequence of the
tidal compression.

\begin{figure}[htbp]
    \centering\plotone{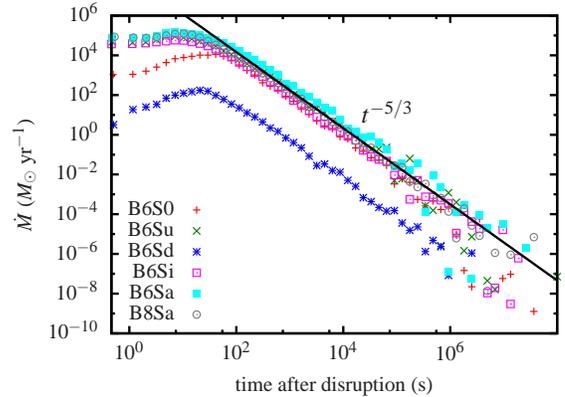}
    \caption{Mass accretion rates computed from the estimated fallback times
    of the fluid elements. The solid line represents the $t^{-5/3}$
    falloff predicted in \citet{Rees:1988bf, Phinney:1989}.}
    \label{fig:fallback-times}
\end{figure}

The late-time behavior consists of material raining back onto the
\bh{} with a characteristic rate of accretion $\dot M \propto
t^{-5/3}$~\citep{Rees:1988bf, Phinney:1989}. This rate is
maintained as long as $dM/d\varepsilon_{\text{kin}}$,
with $\varepsilon_{\text{kin}}$ being the specific kinetic energy, is
approximately constant~\citep{Lodato:2008fr}.  We
follow~\cite{Rosswog:2008ie} and compute the fallback time for each
fluid element at a time long after disruption, based on data available
at the end of the simulation. At this time, hydrodynamical
interactions between the fluid elements are small, and each fluid
element moves on an almost geodesic
orbit. Figure~\ref{fig:fallback-times} shows the results of these
calculations. Clearly we recover the expected $t^{-5/3}$ behavior
for return times longer than $t \gtrsim 200\,\mathrm{s}$. Notice that
earliest times in Figure~\ref{fig:fallback-times} are comparable to
the late time accretion rates found during the simulation in
Figure~\ref{fig:prompt-accretion}.

\section{Electromagnetic Signatures}\label{sec:radiation}

Ultra-close (i.e.  $R_{\rm p} \sim$ few $R_{\rm g}$) tidal disruptions
are violent events.  The conversion of even a small fraction of
gravitational energy into light would result in a powerful
electromagnetic signature.  In this section we will use the hydrodynamics
results from our simulations to make order of magnitude estimates of
the tidal disruption flare's luminosity, its characteristic photon energy,
and the fallback material's electromagnetic signature as it eventually
forms a slim accretion disk near the \bh{.}

When the \whd{} descends into the potential well of the \bh{,} with
the typical dynamical timescale of $\sim 1$~s, it acquires energy in
excess of $\sim10^{53}\,{\rm erg}$. Even if a tiny fraction of
this energy gets converted into radiation on this dynamical timescale,
it would lead to a huge luminosity which would be noticeable at cosmological
distances.  Furthermore, since the \whd{} undergoes a substantial
acceleration in the deep gravitational potential of the \bh{,} the
streams of matter from the disruption will approach speeds $V_{\rm p}
\simeq (G\,M_{\rm h}/R_{\rm p})^{1/2} \simeq c\,(\beta\,R_{\rm
  g}/R_{\rm t})^{1/2} \simeq 0.4\,\beta_6\,c$, comparable to the
speed of
light. The streams of tidal debris later collide and transform a
portion of their kinetic energy into thermal energy.  Being heated
by this process in excess of $10^9$~K,
the dense hot plasma quickly produces photons
\citep{Krolik2011}. These photons, however, are effectively trapped by
the debris, capping the luminosity around the Eddington luminosity
\citep{Eddington1926}
\begin{equation}
  L_{\rm Edd}=\frac{4\,\pi\, c\, G\, M_{\rm h} m_{ p}}{\sigma_T}
  =1.4 \times10^{41}\left(\frac{M_{\rm h}}{10^3\,M_\odot}\right){\rm erg~s}^{-1},
\end{equation}
where $m_p$ is the proton mass and $\sigma_T$ is the Thomson cross-section.
Notice that the equilibrium between the gravitational and spherical radiation
fields is reached at a luminosity $L=2L_{\rm Edd}=2.8 \times 10^{41}{\rm
  erg~s}^{-1}$ due to composition effects.  The system can by far exceed $L_{\rm Edd}$
in the presence of a collimated outflow, especially in the first several minutes after the disruption.
We elaborate on this tidal disruption phenomenon in a subsequent paper (Shcherbakov et al. 2012, in prep).
In the present manuscript we limit ourselves to electromagnetic signatures at late times of months to years.

When the accretion rate falls below the Eddington rate $\dot{M}_{\rm
  Edd}$, the accretion becomes radiatively efficient, radiating mostly
in X-rays with luminosity $\sim L_{\rm Edd}$. Assuming an efficiency
of $\epsilon=0.1$, this luminosity translates into an accretion rate
\begin{equation}\label{Mdot_edd}
\dot{M}_{\rm Edd}=\frac{L_{\rm Edd}}{0.1c^2}=2.3\times 10^{-5}\left(\frac{M_{\rm h}}{10^3\,M_\odot}\right)M_\odot~{yr}^{-1}.
\end{equation}
The time elapsed until the accretion rate drops to $\dot{M}_{\rm Edd}$
can be estimated based on the fallback times reported in
\S~\ref{sec:results}.  We have demonstrated that the famous $t^{-5/3}$
fallback law \citep{Rees:1988bf} is also present in our ultra-close
encounters.  From Figure~\ref{fig:fallback-times}, we observe that
the accretion rate in all but the \Run{b6an6} case drops below
$\dot{M}_{\rm Edd}$ at times of $t_{\rm Edd}\sim1-2$~yrs. Thus the tidal
events we are considering can, in principle, shine at Eddington
luminosities for many years. The actual luminosity could be
substantially sub-Eddington, though, if outflowing debris produced while
the \bh{} swallows matter at super-Eddington rates obscures the accretion
flow. In this section we will first estimate the disk emission within a
slim disk model and then consider the added effects of obscuration by the
outflowing debris.

\subsection{Slim Disk Model}
For our study we employ a slim disk model, developed and summarized in
\citet{Abramowicz1988,Lasota1994,Kawaguchi2003,Sadowski2009}, of a
simple one-zone vertically-integrated dynamical model.
We compute modified black-body bremsstrahlung emission from the disk.
For our radiation estimates, we assume that the angular momentum of
the marginally bound debris accreting at late times, and thus the
angular momentum of the disk, are aligned with the orbital angular
momentum of the incoming \whd{.} The disk model will then, in principle, be
misaligned with the spin of the \bh{.} We presume that, despite the misalignment,
the slim disk extends down to \isco{} radius computed for the aligned case.
In our situation it is challenging to find a more reasonable set of
assumptions for a geometrically thick slim disk. The ultimate answers
require performing full numerical relativity simulations of the
fallback regime, a task requiring modeling dynamical timescales $10^6$
times longer than those addressed in the present study.

We take our heavy white dwarf to consist of approximately $50\%$ carbon
and $50\%$ oxygen as suggested by stellar evolution computations by \citet{Mazzitelli:1986re}.
The same heavy white dwarf composition was assumed by both \citet{Rosswog:2008ie} and \citet{Krolik2011}.
Thus there is $1/14$ nuclei and $1/2$ electrons
per one nucleon with mass $\simeq m_p$.  In terms of nucleon density, the ion
and electron densities are
\begin{equation}
n_{\rm ion}=\frac1{14}n_{\rm nuc} \quad \text{and}\quad \quad n_e=\frac12n_{\rm nuc},
\end{equation}
respectively. The corresponding non-relativistic heat capacity of neutral
and fully-ionized gas per single nucleon of temperature $T$ are
\begin{equation}
c_{\rm neutral}=\frac3{28} k_B T \quad \text{and}\quad c_{\rm ion}=\frac67 k_B T,
\end{equation}
respectively.  The energy of full ionization is
$E_{CNO}=<Z^2>13.6\,eV=680\,eV$ for the assumed composition, where
$Z\approx7$ is the charge of the nucleus.  A CNO \whd{} is thus fully
ionized at $T>T_{\rm ion}\sim5\times10^6{\rm K}$. The molecular mass of
the fully-ionized plasma is $\mu=1.75$, and its density is $\rho=m_p
n_{\rm nuc}$.

A fully self-consistent model of a slim disk requires an estimate of
the disk scale-height $H$. For a disk in the super-Eddington regime,
$H/r=1$~\citep{Abramowicz1988,Kawaguchi2003,Strubbe2009}.  In our simulations, circularization of
debris happens within a distance $30\,R_{\rm g}$ from the \bh{.} We therefore
consider a truncated disk located between the \isco{} radius $r_{\rm
  ISCO}$ and the maximum radius $r_{\max} =30\,R_{\rm g}$. The viscosity parameter
is taken to be $\alpha=0.1$. We solve the equations for the central temperature
as detailed below, taking into account the effects of viscous heating,
emission, and the finite heat capacity of matter. For simplicity, we  follow
\citet{Shakura1973} and set the viscous stress at the \isco{} to
zero. The energy release per unit disk area is then
\begin{equation}\label{eqn:flux_plus}
F_+=\frac{3r_{\rm g}\dot{M}c^2}{8\pi r^3}\left[1-\left(\frac{r_{\rm ISCO}}{r}\right)^{1/2}\right].
\end{equation}
The accretion rate is expressed as
\begin{equation}\label{eqn:mass_conserved}
\dot{M}=4\pi r H \rho_c v_r,
\end{equation}
where $\rho_c$ is the one-zone density and $v_r>0$ is the inflow velocity.

We adopt a beta disk model where the viscous stress is given by
$t_{r\phi}=\alpha p_{\rm g} $ with $p_{\rm g}$ being the gas pressure~\citep{Sakimoto1981}.
A beta disk model is one of three options. In the standard
\citet{Shakura1973} model, $t_{\phi}\propto p_{\rm tot}=p_{\rm
  g}+p_{\rm rad}$, whereas in the ``mean field'' models
$t_{\phi}\propto \sqrt{p_{\rm g}p_{\rm tot}}$ (see
e.g. \citealt{Done2008} for a review). There is no definitive answer
yet about which model better describes stress in a radiation-dominated
accretion disk.  The beta disk model $t_{r\phi}=\alpha p_{\rm g} $ assumption
leads to a radial velocity of plasma
\begin{equation}\label{eqn:v_r}
v_r\approx\alpha c_s=\alpha\sqrt{\frac{k_B T_c}{\mu m_p}}\,,
\end{equation}
where $c_s$ is the isothermal sound velocity. The expression
(\ref{eqn:v_r}) for radial velocity, though not fully
self-consistent, helps to avoid the infinite increase of the
density near the \isco{} inherent to early slim disk models
\citep{Abramowicz1988}.

The accretion disk emits predominantly bremsstrahlung radiation.
Since the scattering cross-section in the disk is much higher than
the absorption cross-section, the emission spectrum is described
by a modified black body spectrum \citep{Rybicki1979,Czerny1987,
Kawaguchi2003}. The photon production rate via bremsstrahlung is
\citep{Rybicki1979,Katz2010}
\begin{equation}
Q=\sqrt{\frac{8}{\pi}}\alpha_f Z^2 n_{\rm ion} n_e
\sigma_T c \sqrt{\frac{m_e c^2}{k_B T_e}}\Lambda_{\rm eff} \,g_{\rm eff} \,{\rm cm}^{-3}{\rm s}^{-1},
\end{equation}
where $\Lambda_{\rm eff}\sim g_{\rm eff}\sim1,$
$\alpha_f\approx0.00729$ is the fine structure constant and $\sigma_T$
is the Thomson cross-section. With the high densities, the production rate
$Q$ is high so the radiation quickly dominates internal energy and pressure.

Let us estimate the equilibrium temperature of this
radiation-dominated flow.  The typical nucleon density in the
simulation varies from $n_{\rm nuc}\sim10^{25}\,{\rm cm}^{-3}$ near the
\bh{,} to a minimum of $n_{\rm min,nuc} \sim 8\times 10^{20}\,{\rm
cm}^{-3}.$ The typical virial temperature is $T_{\rm
ion}\sim10^{11}$~K. As photons are produced, the equilibrium implies
\begin{equation}
\frac67 n_{\rm nuc} k_B T_{\rm ion}=a_{BB} T_{\rm ph}^4\,,
\end{equation}
leading to photon temperatures $T_{\rm ph}\sim3\times10^8$~K.  The
timescale for reaching equilibrium can be estimated as
\begin{equation}\label{eqn:t_eq}
  t_{eq}\sim \frac{a_{BB} T_{\rm ph}^4}{k_B T_{\rm ph}}\frac1{Q(T_{\rm ph})}=9\times
  10^{10} \frac{T_{\rm ion}^{7/8}}{n_{\rm nuc}^{9/8}}\sim 10^{-8}{\rm s},
\end{equation}
which is significantly shorter than the dynamical timescale $t_{\rm
  orb}\sim 1$~s.

We work in the diffusion approximation of high optical depth to find
the surface temperature. The corresponding scattering opacity,
$\tau_{\rm sc}\gg1$, is much larger than the bremsstrahlung absorption
opacity $\tau_{\rm abs}$. The electron scattering opacity is
\begin{equation}
\kappa_{\rm sc}=\frac{\sigma_T n_e}{m_p n_{\rm nuc}}\approx0.2{\rm cm}^2~{\rm g}^{-1},
\end{equation}
whereas the absorption opacity is
\begin{eqnarray}
\kappa_{\rm abs}&=&\frac{(1-e^{-x})e^6 h^2 Z\rho_c}{3c k_B^3 m_e m_p^2 T_s^{7/2} x^3}\sqrt{\frac{2\pi}{k_B m_e}} \nonumber\\
&=&\frac{4.4\times10^{25}{(1-e^{-x})\rho_c}}{T_s^{7/2}x^3}
\end{eqnarray}
where $x=h \nu/(k_B T)$ \citep{Shapiro1986}.  The total emitted power
through one side of the disk with surface temperature $T_s$ is then
\begin{equation}\label{eqn:flux_1}
F_-=\frac{a_{BB} c}4 T_s^4\sqrt{\frac{\bar{\kappa}_{\rm abs}}{\bar{\kappa}_{\rm sc}}},
\end{equation}
where, in a one-zone model, the central density is taken as a proxy for
the surface density. Here $\bar{\kappa}$ denotes mean
opacities. We use here the Rosseland mean opacity \citep{Shapiro1986},
$\bar{\kappa}_{\rm abs}=3.2\times10^{22}Z\,\rho_c\,T_s^{-7/2}$.

The energy flux delivered to the surface is
\begin{equation}\label{eqn:flux_2}
F_{\rm surf}=F_-=\frac{a_{BB}c}3 \frac{T_c^4}{\tau_{\rm sc}}\,.
\end{equation}
To sustain the radiation flow through the surface, this energy flux is equal
to the radiated flux. Thus from
(\ref{eqn:flux_1}) and (\ref{eqn:flux_2}) we obtain that $T_s\propto
T_c^{16/9}$. In a slim disk, cooling does not balance heating, $F_+\ne
F_-$. The residual energy is stored within the accreting material. The
energy balance can then be written as
\begin{equation}\label{eqn:energy_conserved}
Q=F_+-F_-,
\end{equation}
 where the residual energy per unit surface area is given as
\begin{equation}\label{eqn:Q_eq}
Q=\Delta U+A=\frac{4}{3}H a_{BB}T_c^4\left(3\frac{1}{T_c}\frac{dT_c}{dt}-\frac{1}{\rho_c}\frac{d\rho_c}{dt}\right),
\end{equation}
where $dt=-dr/v_r$.  We solve the equations (\ref{eqn:mass_conserved})
and (\ref{eqn:v_r}) for the central density $\rho_c$.  The central
temperature $T_c$ is obtained from equation \ref{eqn:energy_conserved}
with the help of (\ref{eqn:flux_plus}), (\ref{eqn:flux_2}) and
(\ref{eqn:Q_eq}). Given the central density and temperature, we
calculate the surface temperature $T_s$ from (\ref{eqn:flux_1}) and
(\ref{eqn:flux_2}) as well as the corresponding spectrum.

\begin{figure}[htbp]
    \centering\plotone{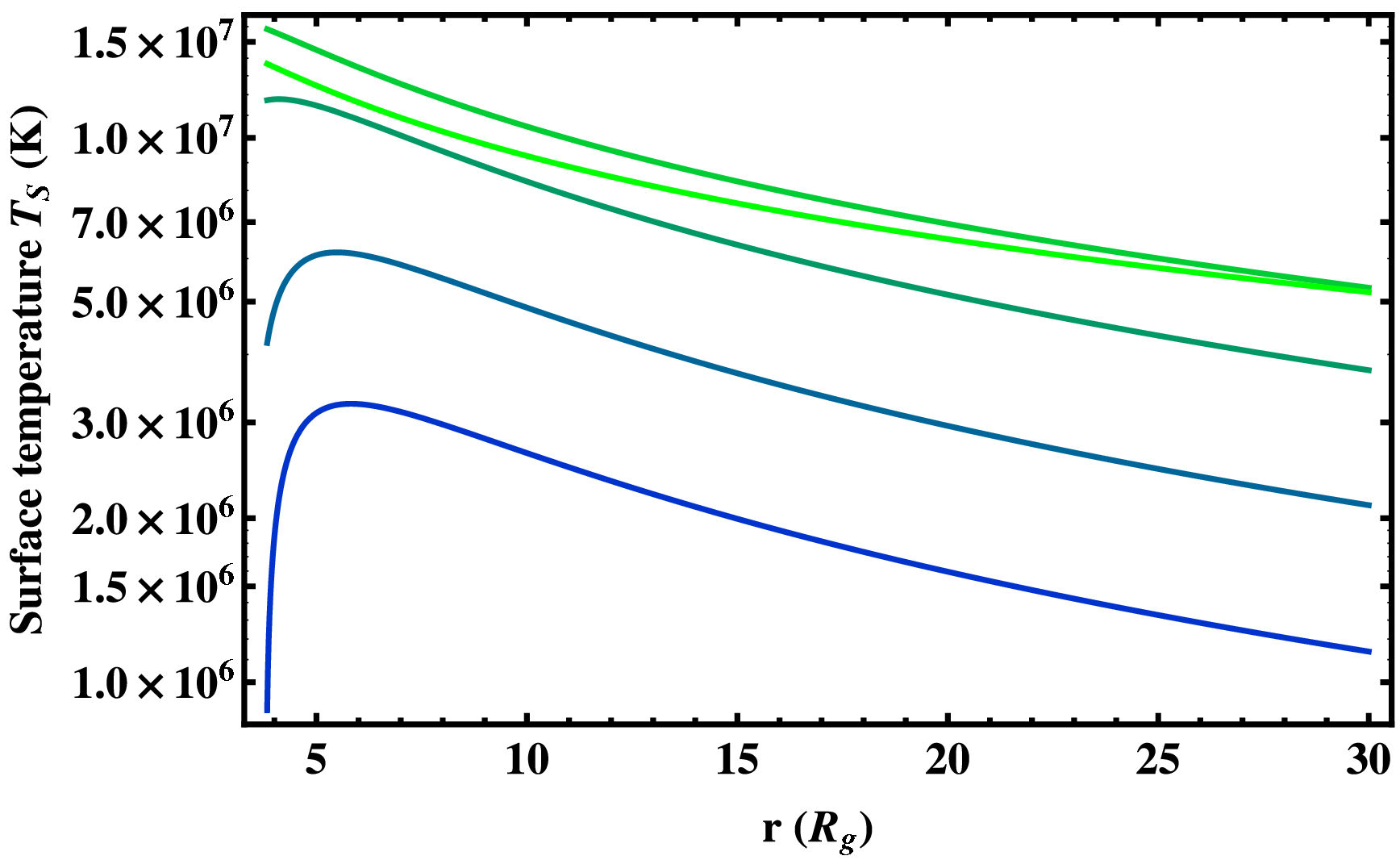}
    \caption{Surface temperature of the slim disk for the case
      \Run{b6ap6} at various times after the disruption. From top to
      bottom: at $t=8$~months, $t=2$~months, $t=3$~years,
      $t=10$~years, and $t=30$~years, respectively. As time progresses
      the accretion rate decreases from $\dot{M}\sim4000\,\dot{M}_{\rm
        Edd}$ to $\dot{M}\sim4\,\dot{M}_{\rm Edd}$. The inner disk
      boundary is chosen to be at $r_{\rm ISCO}=3.83 ~R_{\rm
        g}$.}\label{fig:surface_temperature}
\end{figure}

Figure~\ref{fig:surface_temperature} shows the dependence of surface
temperature $T_s$ on radius $R$ for the accretion rates inferred from
the output of the \Run{b6ap6} simulation. Later times correspond to
lower accretion rates. The temperature $T_s$ is, in general, higher at
higher accretion rates, i.e. earlier times. This is because the energy
release per unit time goes up with accretion rate while the radiating
surface area stays constant. At very low $\dot{M}$, our slim disk
accretion model resembles the standard thin disk model
\citep{Shakura1973} so $T_s\rightarrow0$ at the \isco{.} In contrast,
$T_s$ at the \isco{} saturates at the highest accretion rates.
With little energy release,
$F_+\gg F_-\,$, the energy density is proportional to the accretion rate
$Q=F_+\propto\dot{M}$. In turn, the scattering opacity is also
proportional to the accretion rate $\tau_{sc}\propto n_e\propto
\dot{M}$.  The energy density of radiation is proportional to
$T_c^4$. In addition there is a factor of $\dot{M}$ in both the numerator and
the denominator of the expression (\ref{eqn:flux_2}) for the surface flux.
In the end, $F_{\rm surf}\propto \dot{M}$ in the radiatively inefficient
regime.

\begin{figure}[htbp]
    \centering\plotone{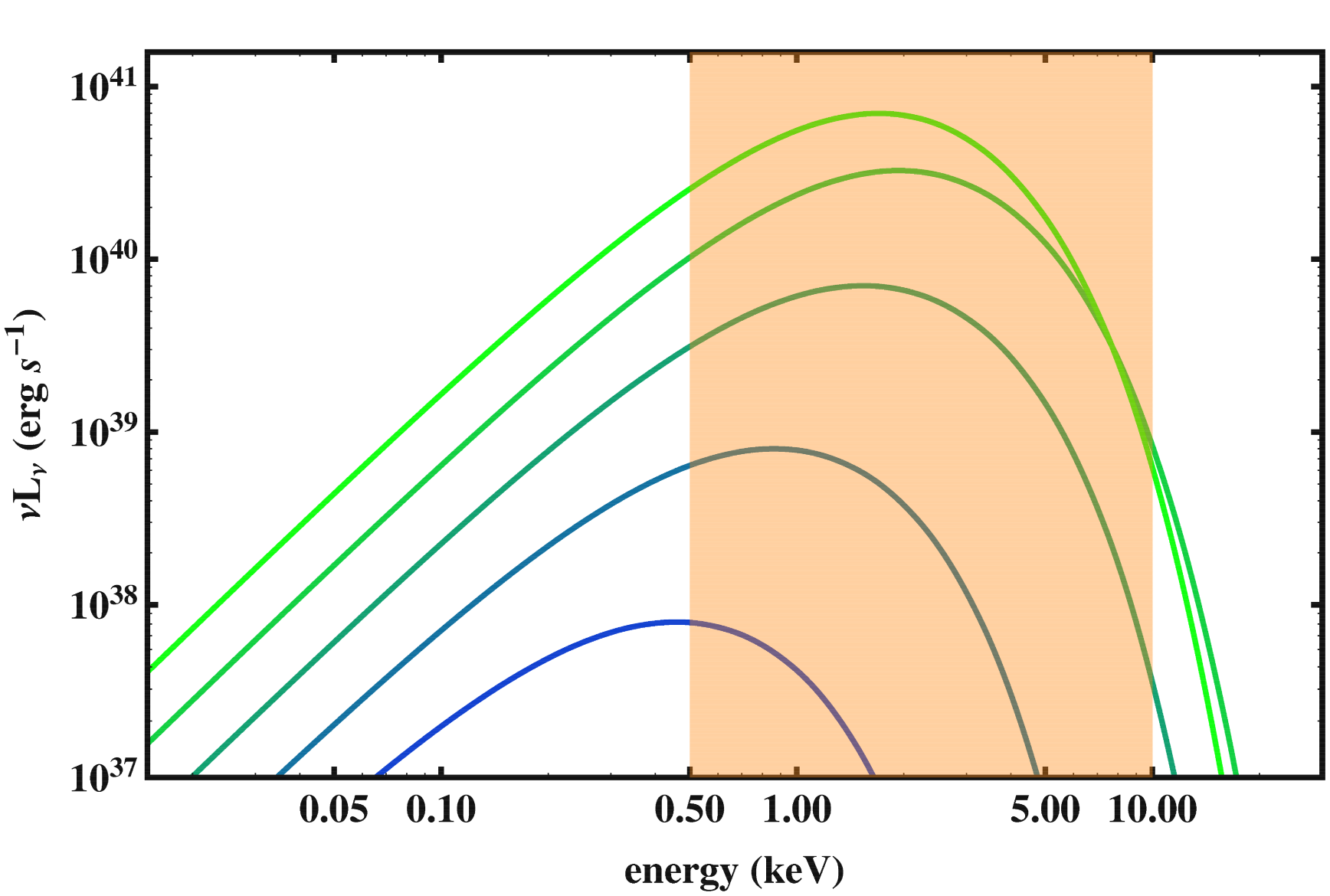}
    \caption{Spectrum for the slim disk formed following the tidal
      disruption at various times. The spectra shown corresponds to
      the case \Run{b6ap6} as in Figure~\ref{fig:surface_temperature}.
      From top to bottom: at $t=2$~months, $t=8$~months, $t=3$~years,
      $t=10$~years, and $t=30$~years, respectively. The inner boundary
      is chosen to be at $r_{\rm ISCO}=3.83R_{\rm g}$. The spectrum
      gets harder with time, but the luminosity decreases.}
    \label{fig:slim_spectrum}
\end{figure}

The corresponding disk spectra are shown in
Figure~\ref{fig:slim_spectrum}. From top to bottom, $\nu L_\nu$
as a function of energy are shown at $2,~8$~months, then $3,~10$ and $30$
years, respectively.  Like $T_s$, the hard tail of the spectrum also
saturates at the highest accretion rates. On the other hand, the
low-energy tail does not saturate at $\dot{M}\sim4000\,\dot{M}_{\rm
  Edd}$. This is because the radiation time is still shorter then or
comparable to the advection timescale at large distances from the BH,
where the low-energy tail is emitted. As the accretion rate drops with
time, the luminosity (the area under the curve) also drops.

\begin{figure}[htbp]
    \centering\plotone{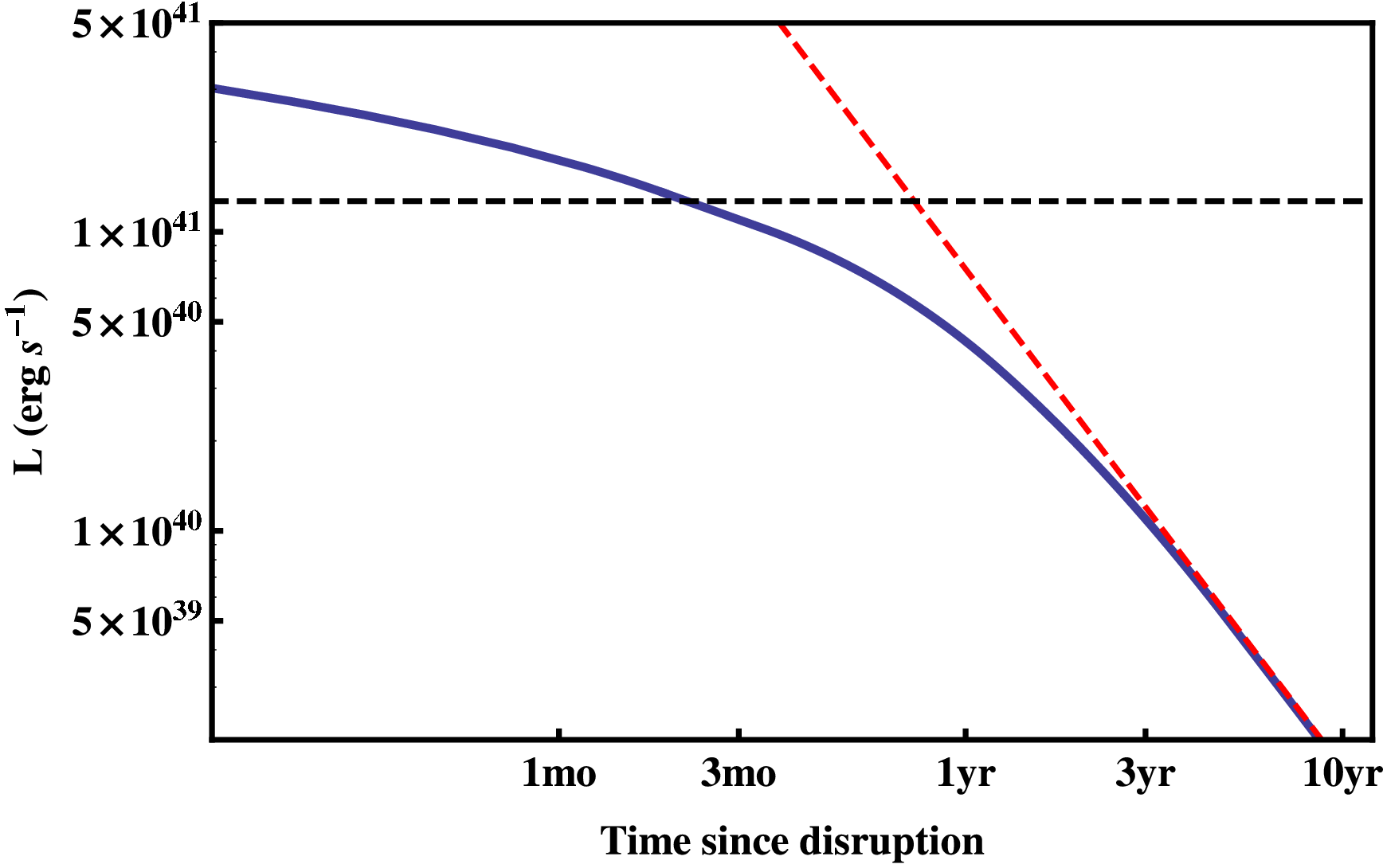}
    \caption{Light curve (solid line) of the slim disk following the
      tidal disruption in the case \Run{b6ap6}. The inner boundary is
      chosen to be at $r_{\rm ISCO}=3.83R_{\rm g}$. The luminosity
      decreases with time, first slowly, then sharply below Eddington
      (horizontal dashed line). The behavior asymptotes to a power-law
      $L =0.05\dot{M}c^2\propto t^{-5/3}$ at low luminosity (inclined
      dashed line), when all thermal energy is radiated on the way to
      \isco{.} The model cannot be trusted if luminosity significantly
      exceeds Eddington value.}\label{fig:slim_lum}
\end{figure}

The integrated luminosity is shown in Figure~\ref{fig:slim_lum} for
the case \Run{b6ap6}. At late times, i.e.  $t\gtrsim2$~years, the
luminosity is proportional to the accretion rate and decreases as
$t^{-5/3}$. Specifically, we estimate a luminosity $L\simeq
0.05\,\dot{M}\,c^2$ for our disk model. Notice that the estimated
luminosity in a fully general relativistic model is $L\simeq
0.091\,\dot{M}\,c^2$ for a thin disk without radiative transfer
effects \citep{Bardeen1972}.  At early times $t\lesssim2$~years the
luminosity saturates at about the Eddington value $L_{\rm
  Edd}=1.4\times10^{41}{\rm erg~s}^{-1}$. The luminosity at times
$t<2$~months is not reliable, since the slim disk model may not be an
adequate representation of accretion at rates
$\dot{M}>4000\,\dot{M}_{\rm Edd}$. For instance, when the outflow from
the inner disk is present, the luminosity may further increase
\citep{Owocki1997,Shaviv2001,Begelman2006,Dotan2011}.  Such a
computation is, however, beyond the scope of the paper.

We consider now the radiative properties of disks from the other cases
with different penetration parameters $\beta$, spins magnitudes
$a^\star$, and spin orientations. Figure~\ref{fig:slim_param} shows
the disk spectra at $t=1$~yr after the disruption for all the
cases studied. In all cases, we assumed a slim disk with inner
boundary at $r_{\rm ISCO}$ and outer boundary at $30\,R_{\rm g}$.

The spectra in Figure~\ref{fig:slim_param} are different because of
two factors. The first contributing factor is the accretion rate
$\dot{M}$, which we parameterize by the fallback time.  More debris
falls back onto the BH in the \Run{b6a6t63p90} case than in the other
cases (see Figure~\ref{fig:fallback-times}). Therefore the spectrum
for the \Run{b6a6t63p90} case is systematically brighter. Similarly,
the lower accretion rate for the \Run{b6a6t90} case makes its spectrum
systematically weaker than the others. We found that one can achieve
coincidence of spectra among all $a^\star=0.6$ cases with a time
shift. That is, at times when the accretion rates $\dot{M}$ are
identical for different simulations, the spectra are also identical.
The spectrum for the \Run{b6a6t63p90} case at about $t=16$~months is
identical to the spectrum for the \Run{b6ap6} case at $t=12$~months
and to the spectrum from the \Run{b6a6t90} case at about $t=9$~months.
Also, more matter is swallowed by the BH initially in the
\Run{b8a6t63p90} case when compared with the \Run{b6a6t63p90} case,
but at later times, less matter that will be eventually accreted is
scattered. Thus the fall back time is lower for the \Run{b8a6t63p90}
case, resulting in a less luminous spectrum. The spectra also vary
because of changes in the \isco{} radius. For instance, the \isco{} is
further from the BH in the \Run{b6a00} case with spin
$a^\star=0$. Since little energy release is taking place inside
$r_{\rm ISCO}=6\,R_{\rm g}$, its spectrum is significantly softer.

\subsection{Obscuration}
\begin{figure}[htbp]
    \centering\plotone{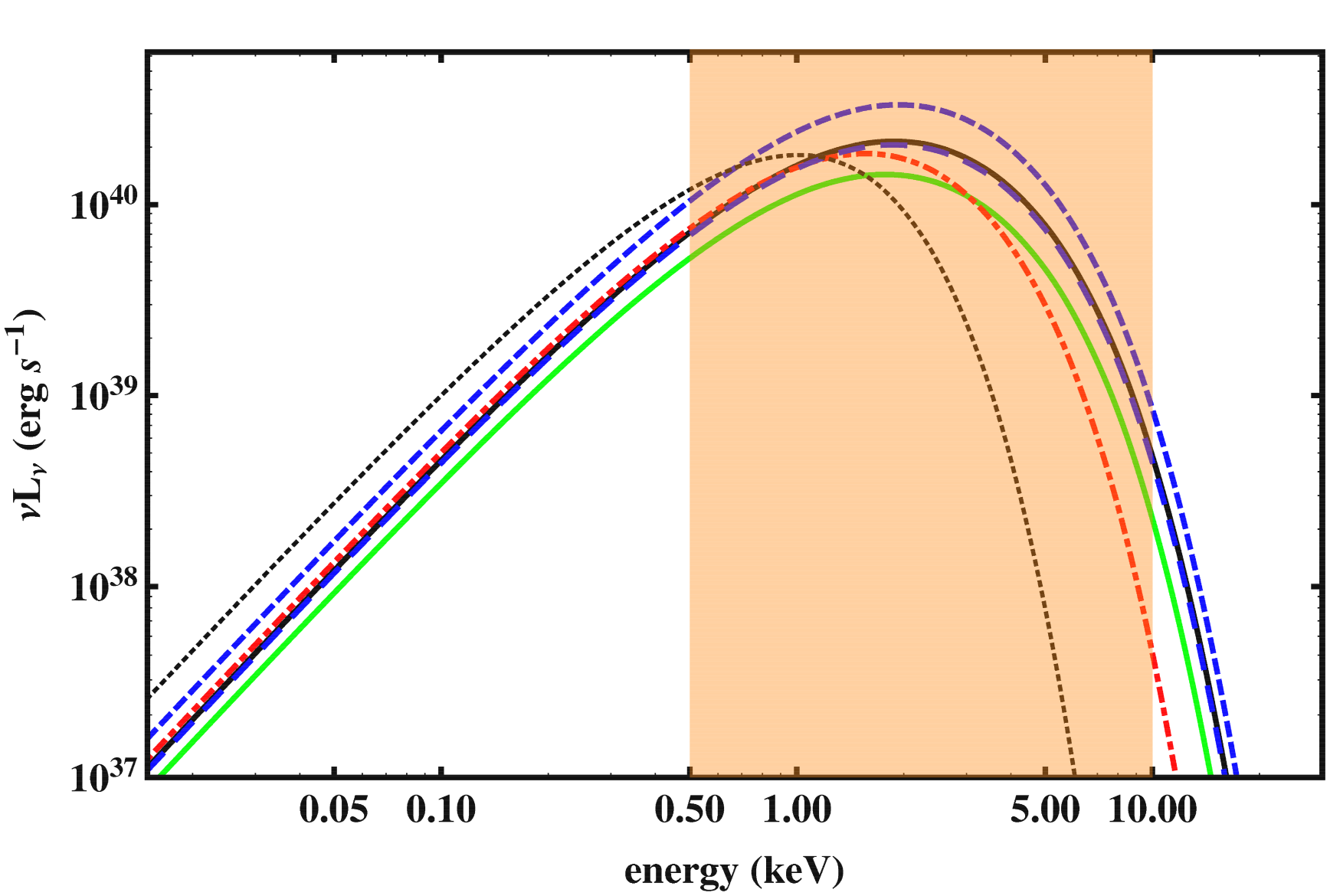}
    \caption{Spectra of our slim disk models at $t=1$~yr: \Run{b6ap6}
      case (top solid/black), \Run{b6a6t90} case (bottom solid/green),
      \Run{b6a6t63p90} case (short-dashed/blue), \Run{b8a6t63p90}
      (long-dashed/blue), \Run{b6a00} case
      (dot-dashed/red). Differences are due to changing spin and
      fallback time. Dotted line corresponds to an edge-on view of a
      slim disk for b6ap6 simulation. The edge-on disk has a softer
      spectrum and a comparable luminosity.}
    \label{fig:slim_param}
\end{figure}

The edge-on view of a slim disk is self-obscured. The spectrum of a
self-obscured disk is not easy to determine. We can approximate the
spectrum by assuming that the thick outer edge at $r=30\,R_{\rm g}$ emits
modified black body radiation the way the top of the slim disk
does. The corresponding edge-on spectrum is included in
Figure~\ref{fig:slim_param} as the black dotted line. It is softer, since only the
material far from the BH contributes. The surface area of the
thick outer edge is large, though, resulting in a raised low-energy tail.

\begin{figure*}[htbp]
\centering\plottwo{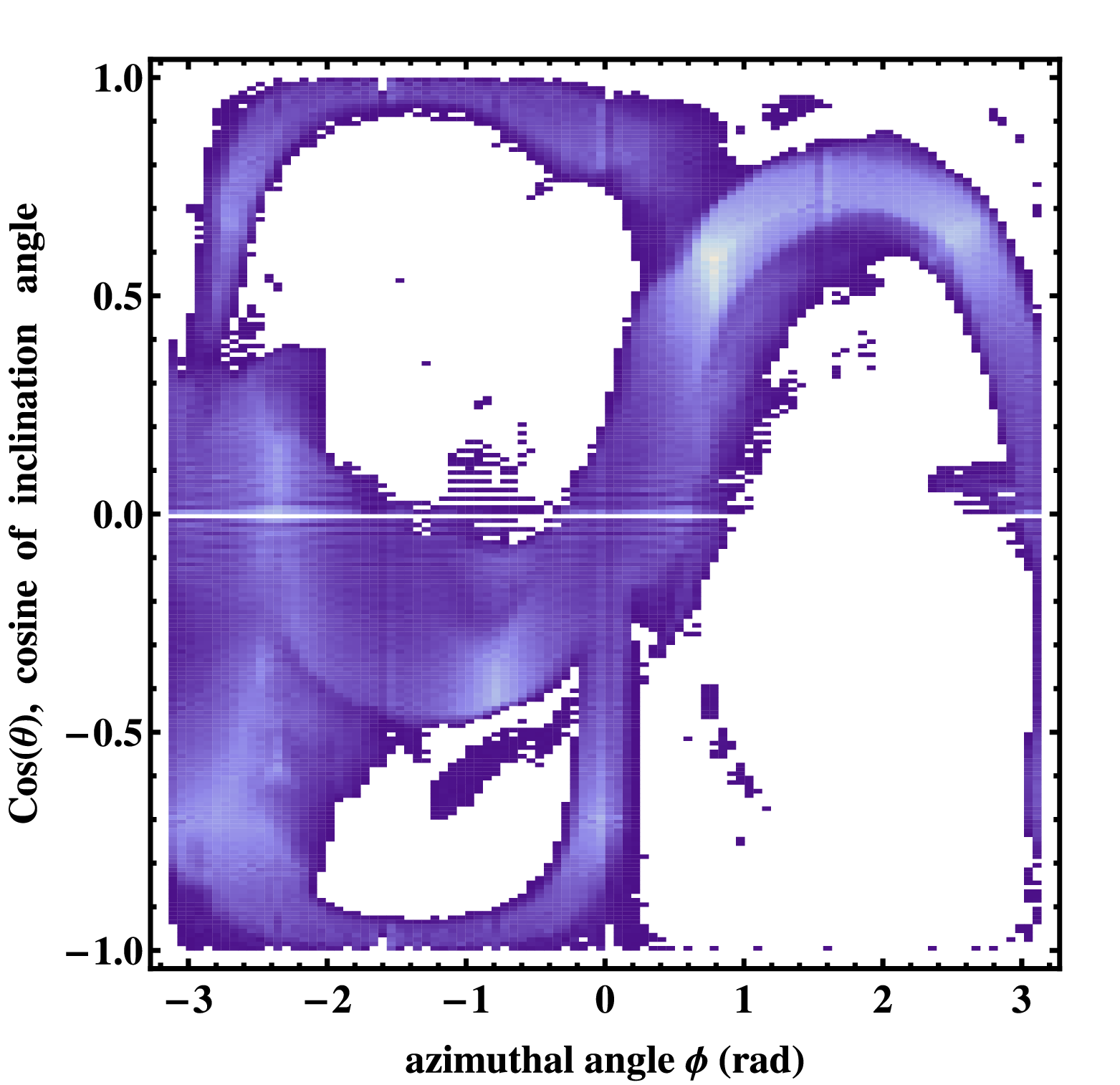}{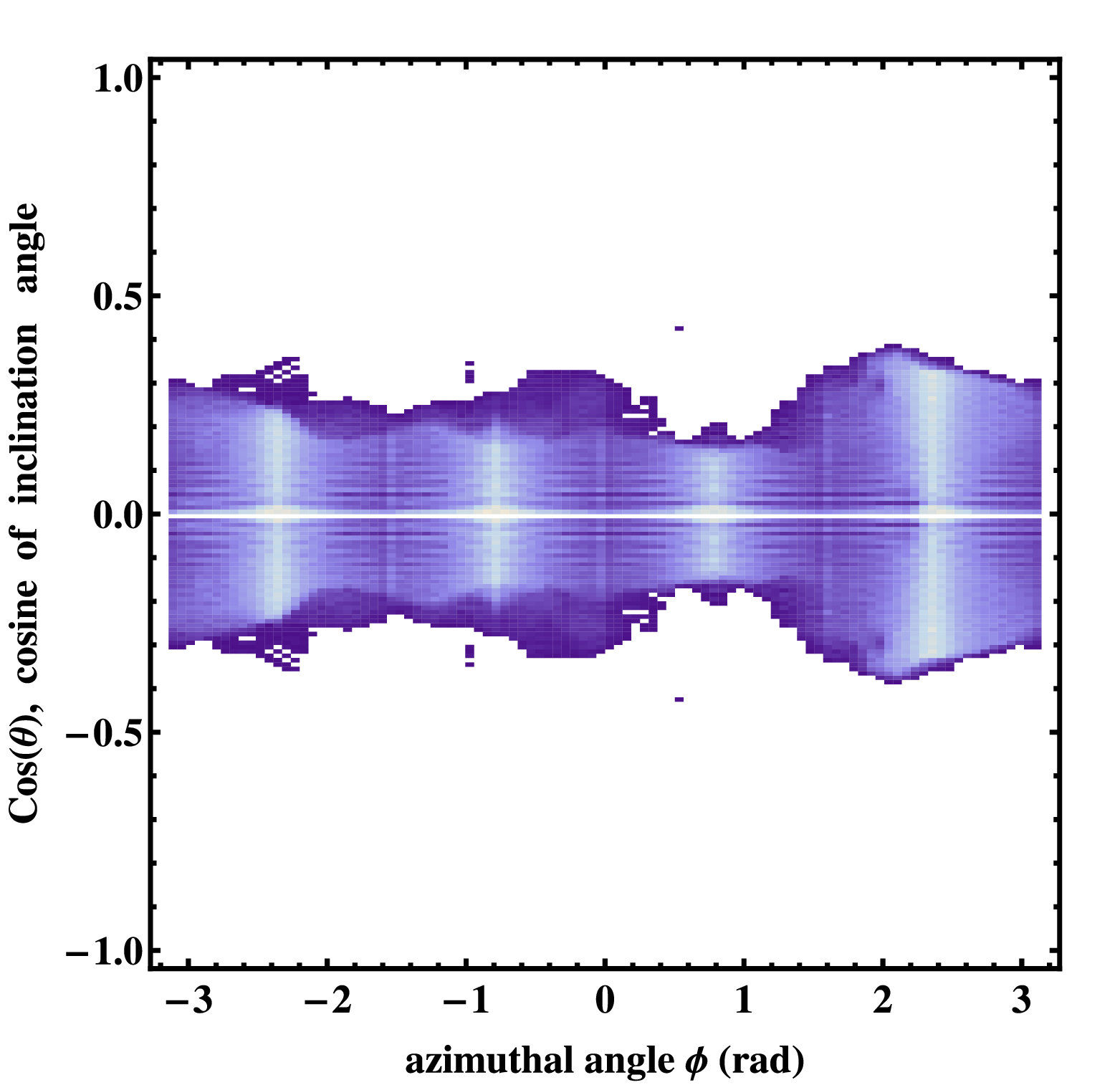}
\caption{Obscuring gas beyond $r=150\,R_{\rm g}$ on the
  $(\cos\theta,\phi)$ plane for the fully misaligned spin case
  \Run{b6a6t90} (\textit{left} panel) and fully aligned spin case
  \Run{b6ap6} (\textit{right} panel). Here $\theta$ is an inclination
  angle of the disruption plane and $\phi$ is the azimuthal angle with
  respect to the \bh{'s} point of view (See Fig.~\ref{fig:setup}).
  \textit{Left:} The gas in
  this case is scattered to all angles, yet there are substantial
  holes through which the inner disk could be visible.
  \textit{Right:} For this case, the disk will be visible in
  directions away from the disruption plane. }
\label{fig:obscuration}
\end{figure*}

Besides exploring the properties of a slim disk, we investigated
whether the emission from the disk is visible through the outflowing
debris surrounding the disk. Figure~\ref{fig:obscuration} shows the
$(\cos\theta,\phi)$ directions in which the debris is scattered in
the ``sky'' from the \bh{} point of view. We considered only the
material outside a sphere of radius $150\,R_{\rm g}$, corresponding
to the boundary between the bound and the unbound matter at time
$t\approx6$~s after the disruption. The unbound matter travels
essentially along straight lines at $t\approx6$~s so
Figure~\ref{fig:obscuration} adequately represents the angular
distribution of matter at late times. White color in
Figure~\ref{fig:obscuration} indicates there is no matter traveling
in that direction whereas all darker colors indicate the presence of
gas.

The optical depth is very high at directions that are not denoted by
white. The inner disk would be completely obscured along those
directions. The left panel in Figure~\ref{fig:obscuration} shows the
distribution of the scattered debris for the \Run{b6a6t90}
case. Notice that although the debris is scattered all over the sky,
there are some preferential directions manifest as a sinusoid in the
$(\cos\theta,\phi)$ plane.  On the other hand, in the case \Run{b6ap6}
shown on the right panel, the debris is scattered preferentially in
the disruption plane. Thus the orientation of BH spin determines the
direction of debris outflow for ultra-close encounters.

We consider now the situation in which we view a \whd{}-BH system with
an unknown orientation. Recall that we have assumed that the disk at
late times is fully aligned with the initial disruption plane. As
discussed before, the disk can, in principle, be viewed both edge-on or
face-on, although the edge-on disk is self-obscured and has a peculiar
spectrum (see Figure~\ref{fig:slim_param}). The probability the disk
is obscured by the outflowing debris is proportional to the solid
angle subtended by the debris.  This solid angle corresponds to the
normalized surface area on the $(\cos\theta,\phi)$ plane (see
Figure~\ref{fig:obscuration}). There may in principle be four
different types of obscuration: i) edge-on self-obscured disk also
obscured by outflowing debris, ii) edge-on self-obscured disk
unobscured by outflowing debris, iii) face-on disk obscured by
outflowing debris, and iv) face-on completely unobscured disk.

\begin{figure*}[htbp]
\centering\plottwo{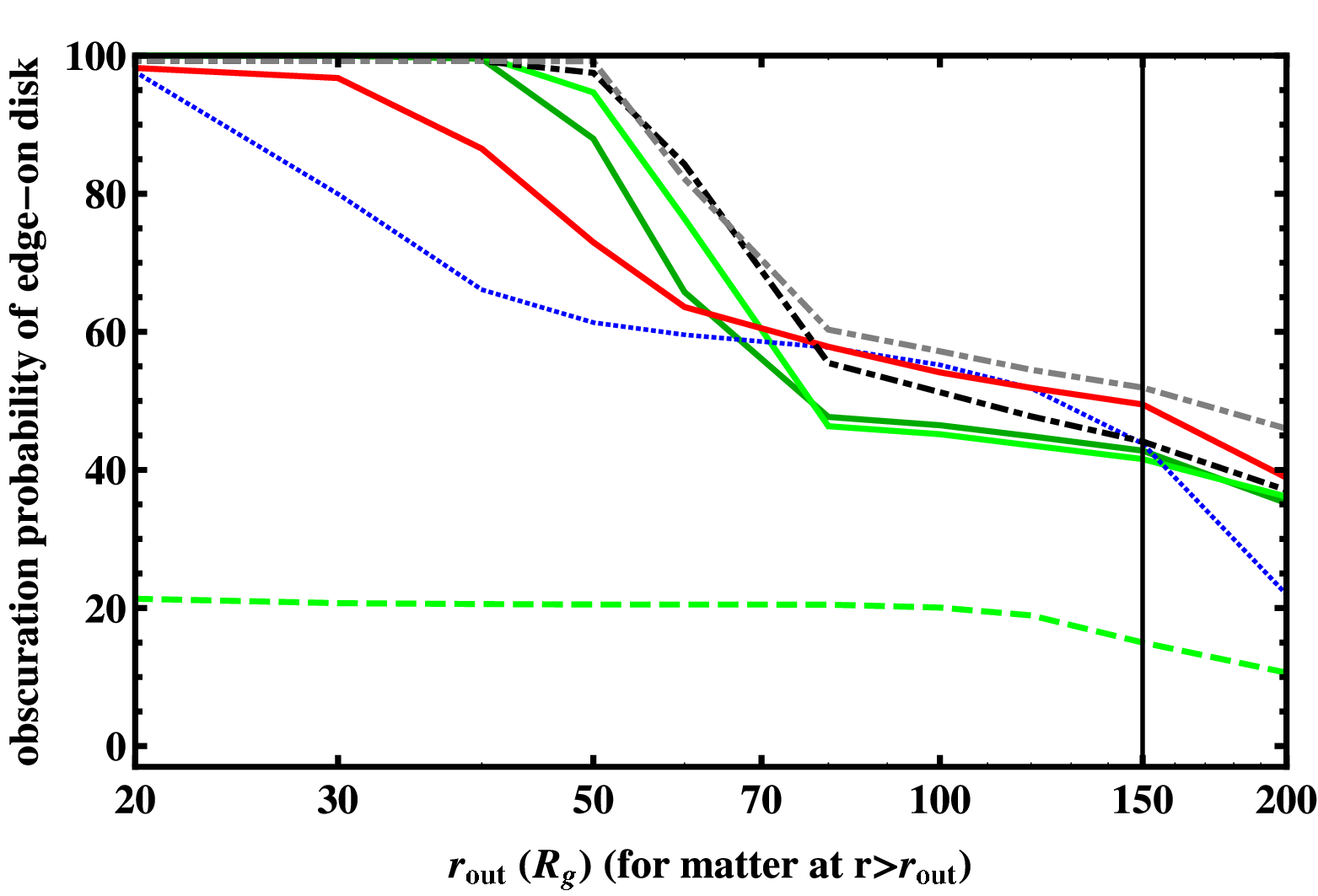}{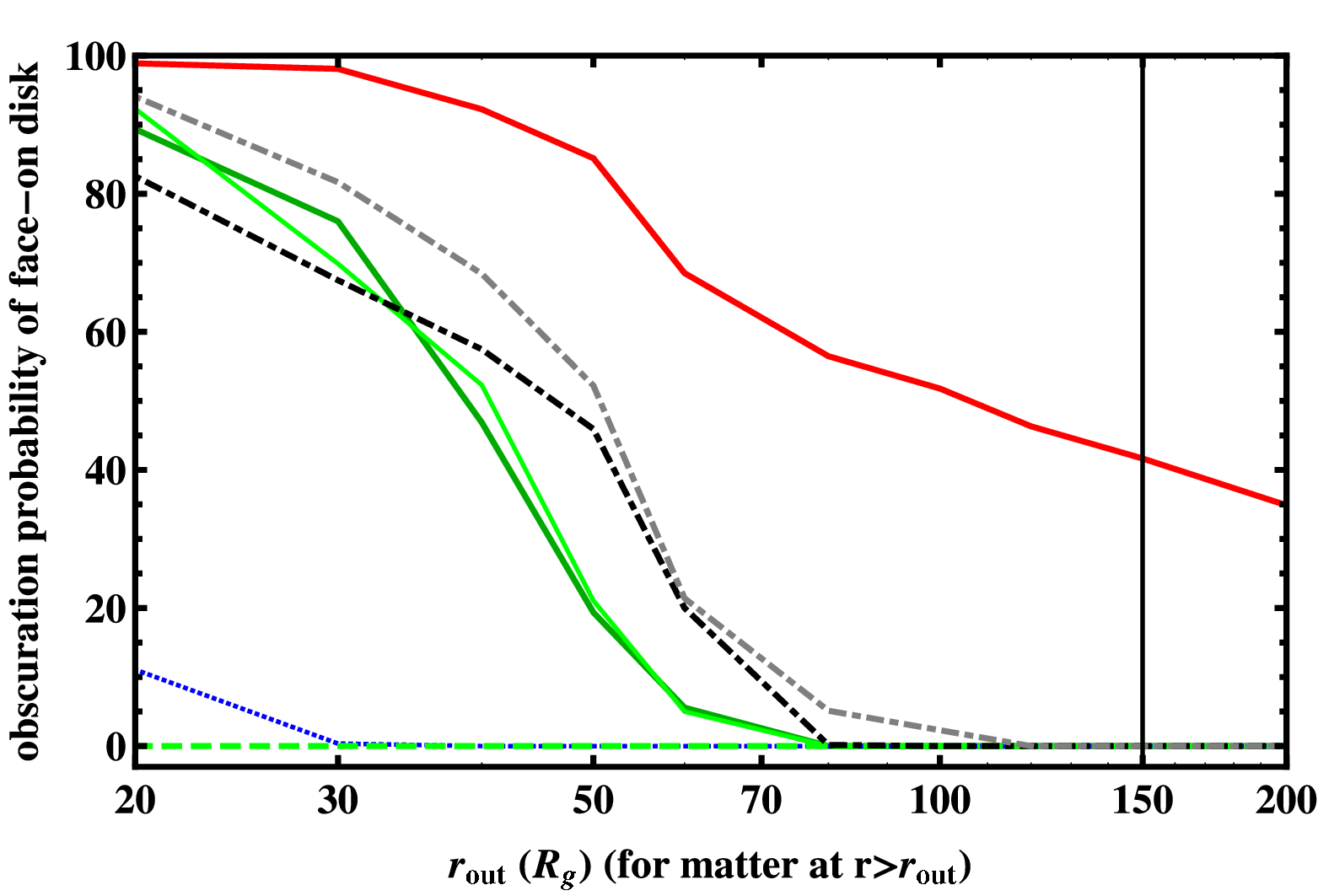}
\caption{Obscuration by debris as a function of the debris beyond
  a radius $r_\mathrm{out}$ from the \bh{.} The obscuration probability is the
  probability that the inner disk is not seen from a random viewing
  angle.  \textit{Left:} obscuration probability for an edge-on disk
  with inclination angles above $45^\circ$ (assuming aligned
  disk). \textit{Right:} obscuration probability for a face-on disk
  with inclination angles below $45^\circ$. The boundary
  of bound/unbound matter is around $r=150\,R_{\rm g}$ at $t=6$~s
  after disruption. The unbound matter moves approximately
  radially so the probabilities at $r_{\rm out}=150R_{\rm g}$
  (denoted by a vertical line) correspond to obscuration probabilities
  days to years later.  Lines types correspond to case \Run{b6a6t90}
  (upper solid red), \Run{b6ap6} (two lower solid green lines: the
  darker ones are computed at earlier time $t=5.4$~s), \Run{b6a00}
  (blue dotted), \Run{b6an6} (green dashed), \Run{b6a6t63p90} (dark
  dot-dashed), and \Run{b8a6t63p90} (light dot-dashed).  The
  obscuration probability of an edge-on disk is about $\approx50\%$
  regardless of spin alignment. Only the \Run{b6an6} case, with very
  little debris, exhibits little edge-on obscuration. In turn, the
  obscuration probability of a face-on disk depends on
  alignment. The fully misaligned \Run{b6a6t90} case is obscured at
  $50\%$ for nearly face-on disk orientations whereas the
  partially misaligned and aligned cases exhibit no obscuration of a
  face-on disk.}\label{fig:cover_fraction}
\end{figure*}

When the height of the photosphere equals the disk height $H=r$
(inclination $\theta=45^\circ$), the probability to encounter a
self-obscured edge-on disk is $p_{\rm edge}=71\%$.  The obscuration
probability by outflowing debris is depicted in
Figure~\ref{fig:cover_fraction} for both edge-on and face-on
disks. The panels show the probability as a function of debris
location at $t=6$~s after the disruption. At that time, the boundary
between the bound and the unbound matter is located at $150\,R_{\rm
  g}$. Thus, this obscuration probability is also the obscuration
probability $p_{\rm obs}$ for days to years later. The obscuration
probability $p_{\rm obs|edge}$ of an edge-on disk can be inferred from
the left panel. Since most debris is scattered in the disruption
plane, this probability is relatively high. The values reach $p_{\rm
  obs|edge}\approx50\%$, practically independent of spin value, spin
orientation or penetration factor. The obscuration probability of a
face-on disk $p_{\rm obs|face}$ can be inferred from the right
panel. Only the fully misaligned case \Run{b6a6t90} scatters
significant amount of debris perpendicular to the disruption plane,
which leads to relatively high $p_{\rm obs|face}\approx40\%$ for this
simulation. Partially and fully aligned simulations all scatter matter
within $45^\circ$ from the disruption plane, and thus cannot obscure
the face-on disk, so that $p_{\rm obs|face}=0\%$. The total
obscuration probability is
\begin{equation}
p_{\rm obs}=p_{\rm obs|face}\,p_{\rm face} + p_{\rm obs|edge}\,(1-p_{\rm face}).
\end{equation}
The values are $p_{\rm obs}\approx50\%$ for the \Run{b6a6t90} case and
$p_{\rm obs}\approx30\%$ for the \Run{b6ap6}, \Run{b6a6t63p90},
\Run{b8a6t63p90} and \Run{b6a00} cases. The misaligned simulation with
$\beta=8$ yields a marginally higher $p_{\rm obs|edge}=52\%$ compared
to a similar simulation with $\beta=6$ for which $p_{\rm
  obs|edge}=44\%$. As expected, deeper penetration results in wider
scattering of debris. However, surprisingly, partially aligned and
fully aligned simulations with $\beta=6$ and $a^\star=0.6$ yield
almost the same obscuration probabilities.

We have previously stated that if there is debris along the line of
sight, then the inner disk is obscured.  However, the optical depth of
debris decays as the debris expands and eventually becomes
transparent. Let us estimate the time $t_{\rm trans}$ when the optical
depth of X-ray absorption equals unity $\tau_{\rm abs}=1$. The actual
time will of course depend on the line of sight. The debris consists
mostly of neutral carbon and oxygen with high absorption cross-section
$\sigma_{\rm abs} \simeq 2\times 10^{-20}{\rm cm}^{-2}$ within
$0.5-10$~keV photon energy range \citep{Henke1993}.  The typical
outflow velocity of the debris is a substantial fraction of the speed
of light $v_{\rm out}\approx0.15c$. The transparency time is then
\begin{equation}\label{eqn:time_trans}
  t_{\rm trans}=\sqrt{\frac{f_{\rm unb}M_{\rm wd}\sigma_{\rm abs}}{56\pi m_p p_{\rm obs}v_{\rm out}^2}}
\end{equation}
where $f_{\rm unb}$ is the fraction of outbound material.  From our
simulations, we found that $f_{\rm unb} \simeq 22\%$ ($f_{\rm
  unb}\simeq 60\%$) for the \Run{b6a6t90} (\Run{b6ap6}) case.  With
our obscuration probability estimate $p_{\rm obs}\simeq 50\%$ ($p_{\rm
  obs}\simeq30\%$ ) for the \Run{b6a6t90} (\Run{b6ap6}) case, we find
that $t_{\rm trans}\approx2$~yr ($t_{\rm trans}\approx4$~yr). The
transparency time given by equation~(\ref{eqn:time_trans}) is an
underestimate, since marginally unbound matter moves at speeds lower
than the average $v_{\rm out}$.

\section{Gravitational wave emission}\label{sec:gravitational-waves}

In addition to producing the electromagnetic signatures discussed in
previous sections, tidal disruptions events are potential sources of
\gw{} radiation. A disruption event generates a burst of \gw{s}
lasting $t \sim R_{\rm p}/V_{\rm p} \sim (R_{\rm p}^3 /G\, M_{\rm bh}
)^{1/2}$\citep{Kobayashi:2004py,Rosswog:2008ie}. The \gw{} burst
produced will have a characteristic frequency $f \sim(G\,
M_\mathrm{bh} / R_{\rm p}^3)^{1/2}$ and amplitude
$h\sim(G\,M_\mathrm{wd}\,R_{\rm g})/(c^2\, R_{\rm p}\, D)$, with $D$
the distance to the \bh{.}  For the case of a \whd{} disrupted by an
\imbh{} \citep{Rosswog:2008ie},
\begin{eqnarray}
\label{eqn:hfreq}
  f &\sim& 0.78\,\beta^{3/2}
  \left( \frac{R_\mathrm{wd}}{0.86\,R_\oplus} \right)^{-3/2}
  \left( \frac{M_\mathrm{wd}}{1\,M_\odot} \right)^{1/2} \,\mathrm{Hz}\\
  \label{eqn:hh}
  h &\sim& 5.9 \times 10^{-20} \beta \left(\frac{D}{20\,\mathrm{kpc}}\right)^{-1}
  \left( \frac{M_\mathrm{wd}}{1\,M_\odot} \right)^{4/3} \nonumber \\
  & &
  \left( \frac{R_\mathrm{wd}}{0.86\,R_\oplus} \right)^{-1}
  \left( \frac{M_\mathrm{BH}}{10^3\,M_\odot} \right)^{2/3}\,.
\end{eqnarray}
These estimates place the \gw{s} from these events at the
low frequency edge of the advanced LIGO design sensitivity range.  At this
edge, the equivalent strain noise is $\sim 10^{-22}\,\mathrm{Hz}^{-1/2}$
~\citep{LIGO-aLIGODesign-Sensitivity}.

\begin{figure}[htbp]
    \centering\plotone{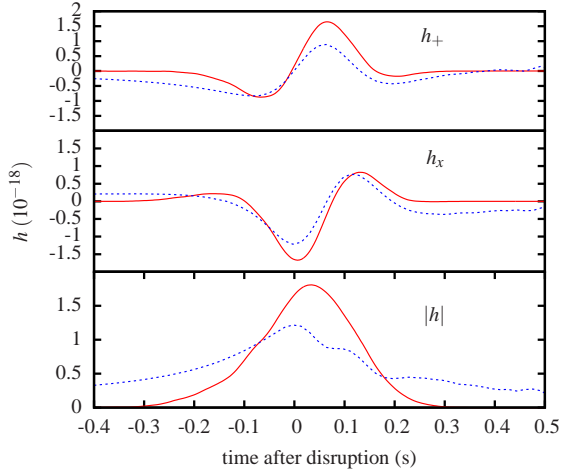}
    \caption{Gravitational wave strain as calculated in full
      general relativity (solid red) and from applying the quadrupole
      formula (dashed blue) for \Run{b6a6t63p90} at a distance of 20 kpc.
      Depicted are the two polarizations, $h_+$ (top) and $h_\times$
      (middle), as well as the amplitude (bottom).}
    \label{fig:gw-quadrupole}
\end{figure}

In Figure~\ref{fig:gw-quadrupole}, we compare the \gw{} strain extracted
from the spacetime (solid red) to that obtained using the quadrupole
formula~\cite{Rosswog:2008ie} applied to the \whd{'s} center of mass
trajectory (dashed blue) for \Run{b6a6t63p90}. While the qualitative 
behavior of the two polarizations, $h_+$ (top) and $h_x$ (middle), are 
similar, the fully nonlinear \gw{} strain differs in both amplitude and phase.

\begin{figure}[htbp]
    \centering\plotone{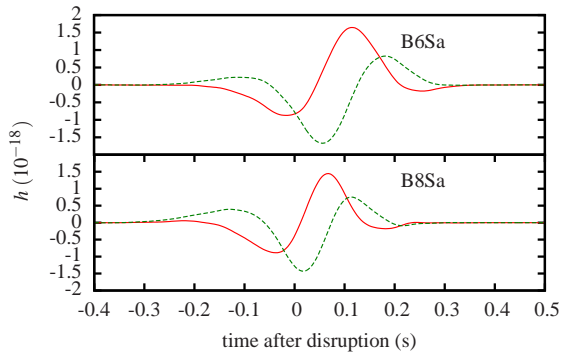}
    \caption{Gravitational wave strain $h$ from disruption events at a
      distance of 20 kpc.  Depicted are the two polarizations, $h_+$
      (solid red line) and $h_\times$ (dashed green line) for cases
      \Run{b6a6t63p90} (top) and \Run{b8a6t63p90} (bottom).}
    \label{fig:gravitational-waves}
\end{figure}

\begin{figure}[htbp]
  \centering\plotone{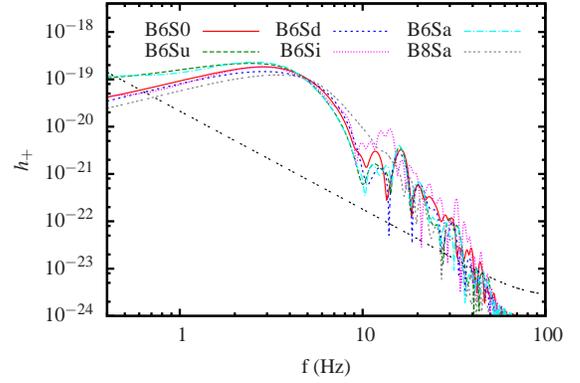}
  \caption{Power spectrum of the $h_+$ polarization of \gw{s} for all
    considered cases. We include the advanced LIGO design sensitivity
    (double-dashed line) for comparison.  All cases indicate a
    characteristic frequency of $~3.2$ Hz.  At low frequencies, $\sim
    1$ Hz, there is a hint of a dependence of the amplitude on the
    spin orientation. }
    \label{fig:gw-spectrum}
\end{figure}

In Figure~\ref{fig:gravitational-waves}, we show the \gw{} strain as a
function of time for disruption events at a distance of 20 kpc.
Depicted are the two polarizations of \gw{s}: $h_+$ (solid red line)
and $h_\times$ (dashed green line) for cases \Run{b6a6t63p90} (top)
and \Run{b8a6t63p90} (bottom).  Time $t=0$ denotes approximately the
time when the disruption takes place. Evident is the burst-like nature
of the \gw{s}. Notice that both cases exhibit comparable burst
duration and strain amplitude. This is expected since the only
difference between \Run{b6a6t63p90} and \Run{b8a6t63p90} is their
value of $\beta$. Although we have stated that case \Run{b6a6t63p90}
has $\beta = 6$ and \Run{b8a6t63p90} a $\beta=8$, from
Table~\ref{tab:accreted-fraction}, we have that the actual penetration
factor for both cases is $\beta^* \simeq 10$.  Interestingly, the
other cases also showed comparable burst durations, $\sim 4$s, suggesting
a weak dependence with the orientation of the spin and penetration
factor.

Figure~\ref{fig:gw-spectrum} shows the power spectrum for all
the cases investigated where we assume that the events took place at
a distance of 20~kpc.  All cases indicate a characteristic frequency
of $\sim3.2$ Hz. This is about an order of magnitude smaller that the
estimated frequency using (\ref{eqn:hfreq}) with $\beta^* \simeq 9$,
the average real penetration factor (see
Table~\ref{tab:accreted-fraction}). This should not be too surprising
since rough estimates using (\ref{eqn:hfreq}) and (\ref{eqn:hh}) do
not take into account, among other things, spin effects. Similarly,
Figure~\ref{fig:gw-spectrum} shows characteristic strain amplitudes of
$\sim 10^{-18}$, an order of magnitude higher than the estimate using
(\ref{eqn:hh}) with $\beta^* \simeq 9$.

At low frequencies, $\sim 1$ Hz, we note that the \Run{b6a6t63p90} and
\Run{b6ap6} cases in particular have higher strain amplitude than the
rest.  Interestingly, these are the same cases which, as seen in
Fig.~\ref{fig:prompt-accretion}, exhibit significantly
lower prompt accretion in the first second after disruption and higher
accretion in the $\sim 1-2$~s after disruption.
The third case with higher prompt accretion in the $\sim 1-2$~s interval,
\Run{b8a6t63p90}, however, exhibits the lowest low-frequency tail.
This suggests that relativistic effects create an entangled parameter space
involving spin and penetration factor during the disruption phase.  Given that
the actual penetration factors in all the cases were
comparable, $7.6 \le \beta^* \le 11$, further study is required to
explore this parameter space.

Synergistic observations of electromagnetic emission from the prompt
accretion discussed in Section~\ref{sec:radiation}, together with
\gw{} detections, may provide ways to measuring the spin of the \bh{}
since in both instances the emission seems to depend, although not
strongly, on the spin orientation. We are currently expanding the
parameter space covered by our simulations to investigate this
dependence in more detail. Multi-messenger observations will be, of
course, only possible if the prompt electromagnetic emission is not
obscured by the debris surrounding the \bh{.}  If such obscuration prevents
observing prompt accretion, the detection of a \gw{} burst would, in
that case, still serve as an identifying precursor to the observations
of the late fallback electromagnetic emission.

\section{Rates and Observational Prospects}\label{sec:observations}
\subsection{Tidal Disruption Rates}
The disruptions of stars by \bh{s} are rare events. More studied are tidal disruptions of stars by SMBHs in galactic centers.
Their rate $10^{-5}{\rm yr}^{-1}$ per galaxy was predicted theoretically
by \citep{Magorrian:1999fh,Wang:2004hg} and found consistent with observations by ROSAT \citep{Donley:2002dj},
XMM-Newton \citep{Esquej2006}, and GALEX \citep{Gezari:2010pr}. The volume density of SMBHs is about \citep{Tundo:2007yu}
\begin{equation}
n_{SMBH}=10^{-2}{\rm Mpc}^{-3},
\end{equation} which leads to tidal disruption rate per unit volume
\begin{equation}
R_{SMBH}=10^{2}{\rm yr}^{-1}{\rm Gpc}^{-3}.
\end{equation}

The rate of disruption of stars in globular cluster (GC) is much less certain. \citet{Baumgardt2004} predicts that, for a GC with $10^3M_\sun$ \bh{},
the optimistic disruption rate of stars is $10^{-7}{\rm yr}^{-1}$ per GC. According to \citet{McLaughlin:1999ty}, the mean GC mass is
$M_{GC}=2.4\cdot10^5M_\sun$, which is larger than clusters considered in \citet{Baumgardt2004}. For a largest cluster simulated by \citet{Baumgardt2004} with $N=131,072$
stars they find $15\%$ of disrupted stars to be \whd{s}. Therefore the optimistic rate of \whd{}-\imbh{} tidal disruptions is
\begin{equation}
D=1.5\cdot10^{-8}{\rm yr}^{-1}
\end{equation} per cluster. The number of GCs changes dramatically from one galaxy to another \citep{Harris:1991hj}.
However, GCs have about constant formation efficiency $\epsilon_{cl}=0.0026$ \citep{McLaughlin:1999ty},
which is the ratio of GC mass to the initial mass of gas in a galaxy.
We assume half of baryonic matter in galaxies, take the Hubble constant $H_0=70.4{\rm km~s}^{-1}{\rm Mpc}^{-1}$ and current baryonic content $\Omega_b=0.00456$ \citep{Komatsu:2011de}
and find the number density of GCs as
\begin{equation}
n_{GC}=0.5\frac{3H_0^2}{8\pi G}\frac{\Omega_b \epsilon_{cl}}{M_{GC}}=34{\rm Mpc}^{-3}.
\end{equation} This density overpredicts the number of GCs in the Milky Way, but is consistent with the overpopulated with GCs elliptical galaxies such as M87.
It is not known, what percentage of GCs contains an \imbh{.} For an optimistic estimates, we take one $10^3M_\sun$ \imbh{} per cluster.
Then, the disruption rate per unit volume is
\begin{equation}\label{eqn:rate_est}
R_{WD-IMBH}=500{\rm yr}^{-1}{\rm Gpc}^{-3}.
\end{equation} This rate is even higher than $R_{SMBH}$, but the \whd{}-\imbh{} disruptions are much less luminous, so they are harder to find.
Also, the rates of tidal disruptions in GCs are quite uncertain \citep{Baumgardt2004},
so even disproving the rate given by equation (\ref{eqn:rate_est}) does not necessarily disprove the hypothesis that every GC has an IMBH.

Out of all tidal disruptions, we focused on ultra-close ones with the ratio of tidal radius to pericenter distance about $\beta=6$.
These constitute a fraction of all disruptions. For a realistic triaxial potential of a GC, chaotic feeding would prevail \citep{Merritt:2004jk}.
Chaotic feeding leads to tidal disruption probability proportional to $R_p$ for objects coming within pericenter distance $R_p$,
as a clear consequence of gravitational focusing. Therefore, tidal disruptions with $\beta>5$ constitute $1/5$ of all events. The correspondent optimistic rate is
\begin{equation}
R_{WD-IMBH,\beta>5}=100{\rm yr}^{-1}{\rm Gpc}^{-3}.
\end{equation}

\subsection{Observations of X-ray Flares}
Considered ultra-close tidal disruptions manifest with a year-long X-ray flare at an Eddington luminosity off the center of the galaxy.
Tidal disruptions by SMBHs rather produce an optical flare \citep{Strubbe2009} in the center of the galaxy.
Let us explore if the optimistic high rates of tidal disruptions are consistent with past observations and if the future observations have a chance of catching such an event.

ROSAT survey probed all sky down to $F_{ROSAT}=4\cdot10^{-13}{\rm erg~s}^{-1}{\rm cm}^{-2}$ flux within $0.1-2.4$~keV band \citep{Voges_ROSAT}.
The peak of X-ray tidal disruption spectrum at $1$~yr (see Figure~\ref{fig:slim_param})
is at about $2$~keV, so that more than half of photons are captured. Taking luminosity $L=4.3\cdot10^{40}{\rm erg~s}^{-1}$
of \Run{b6ap6} simulation at $t=1$~yr we obtain the maximum distance of
\begin{equation}
d_{ROSAT}=25{\rm Mpc}.
\end{equation} Therefore, the expected number of observed \whd{}-\imbh{} disruptions is $N_{ROSAT}=0.015$,
which is consistent with no events. With Chandra, $100$ photons can be obtained over $30$~ks
from the source $200$~Mpc away and over $1$~Ms from the source $1.2$~Gpc away, according to the Portable,
Interactive Multi-Mission Simulator (PIMMS; \citealt{Mukai_PIMMS}).
However, owing to a small field of view of the satellite, the careful choice of field is required to observe at least one event.

In turn, the future mission like Wide Field X-ray Telescope (WFXT) \citep{Conconi:2010hj} can detect a substantial number of events.
It is much more sensitive than ROSAT and observes in the same waveband.
The flux limits of $3\cdot10^{-17}$, $5\cdot10^{-16}$, and $3\cdot10^{-15}$ for deep, medium and wide surveys with correspondent
solid angles $100$, $3000$, and $20,000$ ${\rm deg}^2$ yield maximum event distances $2.5$~Gpc, $600$~Mpc, and $245$~Mpc.
This translates into $76$, $33$, and $16$ \whd{}-\imbh{} disruptions or $5$ times fewer ultra-close disruptions.

Tidal disruptions can also produce powerful prompt super-Eddington X-ray flare with the duration of minutes in addition
to the Eddington-limited X-ray flux on the timescale of $\sim1$~yr. The lightcurve, spectrum and detection prospects
will be discussed in the next paper Shcherbakov et al., (2012, in prep.).

\section{Discussion and Conclusions}\label{sec:conclusions}

Our work addresses disruptions of \whd{s} by \imbh{s}. This is the
first fully general-relativistic study that includes both strong
gravity (the metric of the spinning \bh{)} and dynamical gravity effects (\gw{s}).
We focused on ultra-close disruptions, where the periapsis
radius is comparable to the \bh{} gravitational radius. Our study
presents for the first time results on the influence of the \bh{} spin
on the disruption.

Our study made some simplifying assumptions. Instead of employing a
degenerate equation of state for the \whd{}, we used an ideal gas
law. This is, however, expected to have a very minor effect on dynamics.
If the entropy is approximately constant, the evolution of a
degeneracy pressure supported star is identical to that with an ideal
gas pressure. The choice of pressure prescription does not matter until
shocks and turbulence take over.  Similarly, the absence of explicit
radiation pressure does not influence the initial stages of the
disruption. Cold outflowing debris may not have significant radiation
pressure support, and only the fallback disk will be radiation
pressure dominated. Radiation emitted from the inner disk may halt the
infall of the marginally bound debris. However, the disk is
self-obscured near the disruption plane where most of debris is
scattered, and the photon mean free path is quite small. The photons
cannot thus effectively transfer energy from the inner flow to the
outer flow to unbind the debris. Should radiation pressure be included,
the same internal energy contribute a radiation pressure smaller than the
gas pressure, and thus the total pressure in the medium would be lower.

Our study does not include nuclear reactions, believed to be important
for deep \whd{}
disruptions~\citep{Rosswog:2008ie}. Temperature-density tracks of the matter in
our simulations as depicted in Figure~\ref{fig:temp-vs-rho} are
comparable to those found in Figure~14 of \citet{Rosswog:2008ie}. The
tracks suggest that nuclear burning of carbon and oxygen should
occur in the cases we studied. The additional energy gained in C
and O burning into Fe is $1$~MeV per
nucleon~\citep{Dursi:2005jg}. This translates to a gain of energy of
$\sim 0.001\,M\, c^2$. On the other hand, the descent into the
gravitational potential of the \bh{} yields energies of $\sim 0.1\,M\,
c^2$ for a periapsis radius of $R_{\rm p} \sim 10\,R_{\rm g}$. The
dispersion of debris energies after ultra-close disruptions is also
large. Thus, the extra $0.001\,M\, c^2$ energy from the burning will
not likely have a dramatic effect on the debris. The optical
lightcurve may indeed change after an initially cold white dwarf is
heated by nuclear reactions, but the influence on the fallback disk is
expected to be small.

A realistic fallback disk simulation should incorporate magnetic
fields. Magnetic fields give rise to large viscosity and allow the
disk to self-consistently accrete. The assumed parameter $\alpha$ will
be a proxy for viscosity until magnetized simulations can be run for a
sufficiently long time. The presence of magnetic fields may readily
lead to an outflow~\citep{Blandford1982}, which may boost
luminosity far above the Eddington limit or radiation pressure itself
may drive an outflow under certain
circumstances~\citep{Shaviv2001}. The duration of our simulations
($\sim7$~s) was too limited to directly model the fallback.  The temporal
dependence of accretion rate depicted in
Figure~\ref{fig:fallback-times} is a ballistic guess, based on
a confirmed flat distribution of mass over energy. Any outflow or halted
inflow at late times may result in deviations of the fallback law from
$\dot{M}\propto t^{-5/3}$.

We make a distinction in the paper between the face-on and edge-on
disks. In reality, only minor differences between these orientations
may exist. The disk with extreme accretion rate may drive a wind or
have a corona, which might completely obscure the very inner portions
of the disk. In such a case, the spectrum would be similar to an
edge-on spectrum regardless of orientation.  The precise quantitative
predictions of spectra will only be possible with detailed
numerical simulations of the fallback accretion.

Our study should be viewed as a step towards realistic tidal disruptions
of \whd{s} by \imbh{.} Some of the conclusions may hold up despite our
limited understanding of matter fallback and super-Eddington accretion
flows. Our main conclusions are:

\begin{itemize}
  \item For a non-spinning \bh{} and a \bh{} with a spin
  aligned or misaligned with the orbital angular momentum, the
  debris after disruption forms a thick accretion disk.

  \item For misaligned spins, frame-dragging effects scatter
  the debris around the \bh{} and will often obscure the inner region
  from observation.

  \item There is a qualitatively different behavior before and
  after a timescale of $\sim1$~yr, when the accretion rate
  approaches Eddington accretion rate. The accretion flow luminosity
  stays around $L_{\rm Edd}$ before $\sim1$~yr and starts dropping as
  $t^{-5/3}$ afterwards.

  \item The spectrum peaks at soft X-rays. The spectra are
  similar to thin disk spectra at low accretion rates $\dot{M}\ll
  \dot{M}_{\rm Edd}$,

  \item Self-obscuration and obscuration by debris are
  present in the system. Self-obscuration leads to softer spectrum while
  obscuration by debris may make a fallback disk invisible during
  most of the active accretion period.

  \item The \gw{} signal depends weakly on the orientation of the
  spin. The \gw{} burst will be challenging to be detected for
  extragalactic sources.

\end{itemize}

\section{Acknowledgements}\label{sec:acknowledgements}

The results reported in this paper were obtained with the the \maya{}
code of the numerical relativity group at Georgia Tech.  The \maya{}
code solves the Einstein equations of general relativity using the
\BSSNOK{} formulation, and \bh{s} are modeled following the moving
puncture method~\citep{Campanelli:2005dd,Baker:2005vv}.  The source code
is
partially generated using \kranc{}~\citep{Husa2006}. The
\maya{} code works under \cactus{,} with adaptive mesh refinement
provided by \carpet{}~\citep{Schnetter2004}. Hydrodynamics in the
\maya{} code is based on the public version of the \whisky{} code
developed by the European Union Network on Sources of Gravitational
Radiation~\citep{Baiotti:2004wn}.  We construct initial data using the
\twopunctures{} code of~\cite{Ansorg:2004ds}, and use
\ahfinderdirect~\citep{Thornburg:2003sf} to locate the apparent horizon.
Both codes are part of the Einstein Toolkit~\citep{EinsteinToolkit:web}.
Details of the \maya{}
code can be found
in~\citet{Vaishnav:2007nm,Hinder2008b,Herrmann2007b,Herrmann2007a,Bode2008,Bode2009a,Bode:2011ac}.

The authors are thankful to Ramesh Narayan for fruitful discussions.
RVS is supported by NASA Hubble Fellowship grant HST-HF-51298.01.
RH gratefully acknowledges support by the Natural Sciences and Engineering
Council of Canada.
Work supported by NSF grants 0653443, 0855892, 0914553, 0941417, 0903973,
0955825. Computations at Teragrid TG-MCA08X009 and the Georgia Tech
FoRCE cluster.

\bibliographystyle{apj}
\bibliography{BBL/refs}

\end{document}